\documentclass[letterpaper,11pt]{article}
\pdfoutput=1
\usepackage{jheppub}
\usepackage{slashed}
\usepackage{graphicx}
\usepackage{subfigure}
\usepackage{soul}
\usepackage{amsmath}
\usepackage{multirow}
\usepackage{tablefootnote}
\usepackage{hyperref}
\usepackage{epstopdf}
\usepackage{lscape}
\usepackage{url}
\usepackage[normalem]{ulem}

\usepackage[dvipsnames]{xcolor}
\usepackage[utf8]{inputenc}

\newcommand{\beq}{\begin{equation}}
\newcommand{\eeq}{\end{equation}}
\newcommand{\bea}{\begin{eqnarray}}
\newcommand{\eea}{\end{eqnarray}}

\newcommand{\changed}[2]{{\protect\color{red}\sout{#1}}{\protect\color{blue}\uwave{#2}}}

\def\m1{M_1}
\def\m2{M_2}
\def\m3{M_3}

\def\ch10{\tilde \chi^0_1}

\def\to{\rightarrow}

\newcommand{\lsim}{\mathrel{\mathop{\kern 0pt \rlap
  {\raise.2ex\hbox{$<$}}}
  \lower.9ex\hbox{\kern-.190em $\sim$}}}
\newcommand{\gsim}{\mathrel{\mathop{\kern 0pt \rlap
  {\raise.2ex\hbox{$>$}}}
  \lower.9ex\hbox{\kern-.190em $\sim$}}}

\definecolor{pink}{RGB}{255,105,180}

\definecolor{green2}{rgb}{0,0.56,0.32}

\def\figureautorefname~#1\null{Fig.\,#1\null}
\def\tableautorefname~#1\null{Tab.\,#1\null}

\def\equationautorefname~#1\null{Eq.\,(#1)\null}

\title{Charged Higgs Search in 2HDM}

\author[a]{Juxiang Li,}
\author[b]{Huayang Song,}
\author[c]{Shufang Su,}
\author[a]{Wei Su}

\affiliation[a]{School of Science, Shenzhen Campus of Sun Yat-sen University, No. 66, Gongchang Road, \\ Guangming District, Shenzhen, Guangdong 518107, P.R. China }
\affiliation[b]{Particle Theory and Cosmology Group, Center for Theoretical Physics of the Universe, Institute for Basic Science (IBS), Daejeon, 34126, Korea}
\affiliation[c]{Department of Physics, University of Arizona, Tucson, Arizona 85721, USA}

\emailAdd{lijx376@mail2.sysu.edu.cn}
\emailAdd{huayangs@ibs.re.kr}
\emailAdd{suwei26@mail.sysu.edu.cn}
\emailAdd{shufang@arizona.edu}

\abstract{
In this paper, we present a comprehensive study of the collider search limits on the charged Higgses in the  four types of Two Higgs Double Models (2HDM).  In addition to constraints from flavor physics measurements, we include both the LEP charged Higgs search channels, as well as the LHC search results on the light and heavy charged Higgses.  We consider both the conventional charged Higgs search channels of $H^\pm \rightarrow \tau\nu, cs, cb, tb$, and  the latest search results on the exotic decay channels $H^\pm \rightarrow A W^\pm / H W^\pm$.  We find that $H^\pm \rightarrow A W^\pm / H W^\pm$ are complementary to the conventional fermionic channels for $m_{H^\pm} < m_t$.  For heavy $H^\pm$,  $H^\pm \rightarrow A W^\pm / H W^\pm$ extend the reach of $\tan \beta$ beyond that of $H^\pm \to tb$ in the Type-L 2HDM.  We also present the combined reach of all the neutral and charged Higgs searches.  }


\begin{document}
\maketitle
\flushbottom
\tableofcontents
	
\section{Introduction} \label{sec1}
Since the discovery of the 125 GeV Standard Model (SM)-like Higgs boson by the ATLAS~\cite{ATLAS:2012yve} and CMS~\cite{CMS:2012qbp} collaborations at the Large Hadron Collider (LHC), the SM is confirmed to be a self-consistent theory at the electroweak scale. At the same time there are still unsolved puzzles in particle physics, such as the existence of dark matter, the baryon asymmetry of the universe, the strong CP problem, the muon $g-2$ anomaly. These problems strongly motivate new physics beyond the Standard Model (BSM).  

Many of the proposed new physics models contain an extended Higgs sector, among which the Two Higgs Doublet Model (2HDM) is one of the simplest extensions~\cite{Lee:1973iz, Branco:2011iw}. After electroweak symmetry breaking (EWSB), the Higgs sector of the 2HDM consists of five physical Higgs bosons, including two CP-even neutral scalars $h$ and $H$,  a CP-odd Higgs boson $A$, and a pair of charged Higgs bosons $H^{\pm }$~\cite{Branco:2011iw}.  Studies of the existing constraints on the 2HDM charged Higgses have been performed in Ref.~\cite{Arbey:2017gmh}, focusing on the direct searches for charged Higgses at the LHC via conventional channels $H^\pm\rightarrow f f^\prime$, as well as the indirect constraints of flavor physics   for four different types of the 2HDM. However, the exotic channels $H^{\pm} \rightarrow W^{\pm}\phi(\phi=A/H/h)$ have been studied at the LEP (~\cite{Akeroyd:1998dt, Akeroyd:2000xa, ALEPH:2013htx}), the LHC (\cite{Coleppa:2014cca,Kling:2015uba,Bahl:2021str}), as well as the future hadron colliders (\cite{Kling:2018xud,Li:2020hao}). In the current study, we focus on the constraints on the charged Higgses including both the conventional channels and exotic decays of $H^\pm \rightarrow AW/HW$, for four different types of 2HDM.

If the mass of the charged Higgs $m_{H^{\pm }}$ is less than the top quark mass $m_{t}$, the primary production channel of the charged Higgs is through  top decay $t\to H^{\pm }b$. The charged Higgs boson primarily decays into $\tau \nu$ or $cs$, as well as Cabibbo–Kobayashi–Maskawa (CKM) suppressed channel $cb$. For heavier charged Higgs  $m_{H^{\pm }} > m_{t}$, the main production mode is the top quark associated production $pp\to H^{\pm }tb$, with  $H^{\pm } \to tb $ being dominant.  Once kinematically accessible, the decays of $H^{\pm } \to AW/HW$  quickly dominate over the conventional search channels of $H^\pm \rightarrow f f^\prime$.  While the constraints from conventional channels relax, $H^{\pm } \to AW/HW$  provide  alternative channels for the charged Higgs searches that are complementary to the conventional modes~\cite{Coleppa:2014cca,Kling:2015uba,Kling:2018xud,Li:2020hao,Bahl:2021str}.

In this paper, we analyze the constraints on the charged Higgs boson from direct searches at the LHC and LEP, taking into account the latest experimental data.  We also present an up-to-date analysis of the combined flavor constraints on the charged Higgs boson in the 2HDM, based on the latest flavor physics measurements~\cite{HFLAV:2019otj,Banerjee:2024znd,ParticleDataGroup:2024cfk,Chen:2024cqn} of the branching ratios of $B\to \chi_{s} \gamma$, $B_s\to \mu^ + \mu^-$, $B_d\to \mu^ + \mu^-$, $B\to \tau \nu$, $D_s\to \mu \nu$ and $D_s\to \tau \nu$ from the LHC and the $B$ factories.  

The rest of the paper is organized as follows. In \autoref{sec2}, we introduce the theoretical framework of the 2HDM. In \autoref{sec3}, we summarize the latest LHC and LEP searches that are relevant to charged Higgs studies, as well as flavor constraints. In \autoref{sec4}, we present the constraints on the 2HDM parameter spaces of Type-I, Type-II, Type-L and Type-F. We conclude in \autoref{sec5}.


\section{Framework of the 2HDM} \label{sec2}
 The most general Higgs potential of the CP-conserving 2HDM with  a $\mathcal{Z} _{2}$ soft-breaking term is given by~\cite{Branco:2011iw}
\begin{eqnarray}
\label{eq:V_2HDM}
 V&=& m_{11}^2\Phi_1^\dag \Phi_1 + m_{22}^2\Phi_2^\dag \Phi_2 -m_{12}^2(\Phi_1^\dag \Phi_2+ h.c.) + \frac{\lambda_1}{2}(\Phi_1^\dag \Phi_1)^2 + \frac{\lambda_2}{2}(\Phi_2^\dag \Phi_2)^2  \notag \\
 & &+ \lambda_3(\Phi_1^\dag \Phi_1)(\Phi_2^\dag \Phi_2)+\lambda_4(\Phi_1^\dag \Phi_2)(\Phi_2^\dag \Phi_1)+\frac{\lambda_5}{2} \left[(\Phi_1^\dag \Phi_2)^2 + h.c.\right]\, ,
\end{eqnarray}
in which the two complex hypercharge $Y=1/2$, ${\rm SU(2)}_L$ doublets are denoted as $\Phi _{1, 2}$, and all the parameters are real.  After the EWSB, the doublets can be parameterized as
\begin{equation} \label{eq:double}
\Phi _i =\begin{pmatrix}  
  \phi _{i}^{+}  \\  
  \left ( v_{i}+\phi _{i}^{0} + i\varphi _{i }     \right ) /\sqrt{2}
\end{pmatrix}, \ \ i = 1,2,
\end{equation}
where $v_{1,2}$ are the vacuum expectation values of the neutral components which satisfy the condition $\sqrt{v_{1}^{2}+v_{2}^{2} }\equiv v $= 246 GeV.  For the purpose of this paper, the scalar sector of the 2HDM is   parameterized by the physical Higgs mass ($m_{h}$, $m_{H}$ ,$m_{A}$ and $m_{H^{\pm } } $),  the mixing angle between the two CP-even Higgses ($\alpha $), and the ratio of the two vacuum expectation values ($\tan\beta = v_{2} /v_{1}$), as well as a soft $\mathcal{Z} _{2}$ breaking parameter $m_{12}^{2}$.  

After the EWSB, the Higgs mass eigenstates contain a pair of CP-even Higgses $h$ and $H$, a CP-odd Higgs $A$, and a pair of charged Higgses $H^{\pm}$ through:
\begin{eqnarray} \label{eq:mass}
h=-s_{\alpha }\phi _{1}^{0} + c_{\alpha } \phi _{2}^{0} ,&& H= c_{\alpha }\phi _{1}^{0} + s_{\alpha } \phi _{2}^{0},\\
A=-s_{\beta  } \varphi_{1} + c_{\beta } \varphi _{2},&&
H^{\pm } =-s_{\beta  }\phi _{1}^{\pm } + c_{\beta } \phi_{2}^{\pm },
\end{eqnarray}
where we use the shorthand notation $s_{x} = \sin x$ and $c_{x} = \cos x$. In the following, we will identify $h$ as the discovered 125 GeV Higgs. For general parameter choices, the exotic decay channels of the charged Higgs $H^{\pm } \to hW^{\pm }, H^{\pm } \to HW^{\pm }$ and $H^{\pm } \to AW^{\pm } $ can be kinematically accessible. The couplings governing these processes are given by
\begin{eqnarray} \label{eq:Vcoup}
g_{H^{\pm }hW^{\mp} } = \frac{gc_{\beta - \alpha } }{2} \left ( p_{h} - p_{H^{\pm } }  \right )_{\mu}, &&
g_{H^{\pm }HW^{\mp} } = \frac{gs_{\beta - \alpha } }{2} \left ( p_{H} - p_{H^{\pm } }  \right )_{\mu}, \\
g_{H^{\pm }AW^{\mp} } = \frac{g}{2} \left ( p_{A} - p_{H^{\pm } }  \right )_{\mu} ,&&
\end{eqnarray}
where $g$ is the ${\rm SU(2)}_{L}$ coupling constant and $p_i$ are the incoming momentum of the corresponding particle $i$. Higgs coupling measurements at the LHC prefer the alignment limit $s_{\beta - \alpha}=1$. Therefore the coupling of the charged Higgs to the 125 GeV Higgs and the gauge boson is suppressed by $c_{\beta - \alpha }\sim 0$ and $H^{\pm }$ is more likely to decay into the non-SM-like CP-even Higgs and a gauge boson $HW^{\pm}$ or the CP-odd scalar and a gauge boson $AW^{\pm}$.

To avoid flavor-changing neutral currents (FCNC), one usually imposes a $\mathcal{Z} _{2}$ symmetry on the fermion and Higgs sectors,  under which each fermion type only couples to one Higgs doublet. Depending on how fermions couple to $\Phi_{1,2}$, there are four possible types of 2HDMs: Type-I, Type-II, Type-L (lepton-specific) and Type-F (flipped), as shown in \autoref{tb:couping}.
\begin{table}[!ht]
\centering
\begin{tabular}{|c|c|c|c|c|c|c|}
\hline
2HDM &up-type ~ &down-type ~ &lepton ~ & $\xi _{u}$ ~ & $\xi _{d}$ ~ & $\xi _{e}$ ~ \\  
\hline
\cline{2-6}
Type-I& $\Phi _{2}$ ~ & $\Phi _{2}$ ~ & $\Phi _{2}$ ~ & $\cot  \beta$ ~ & $\cot  \beta$ ~ & $\cot  \beta$ ~  \\ 
\hline
\cline{2-6}
Type-II& $\Phi _{2}$ ~ &$ \Phi _{1}$ ~ & $\Phi _{1}$ ~ & $\cot  \beta$ ~ & $-\tan  \beta$ ~ & $-\tan  \beta$ ~ \\ 
\hline
\cline{2-6}
Type-L& $\Phi _{2}$ ~ & $\Phi _{2}$ ~ & $\Phi _{1}$ ~ & $\cot  \beta$ ~ & $\cot  \beta$ ~ & $-\tan  \beta$ ~ \\ 
\hline
\cline{2-6}
Type-F& $\Phi_{2}$ ~ & $\Phi_{1}$ ~ & $\Phi_{2}$ ~ & $\cot  \beta$ ~ & $-\tan  \beta$ ~ & $\cot  \beta$ ~  \\ 
\hline
\cline{2-6}
\end{tabular}
\caption{Types of 2HDM, along with $\xi_{u,d,e}$ as defined  in \autoref{eq:fcouping}. }

\label{tb:couping}
\end{table}

The Yukawa interactions~\cite{Kanemura:2015mxa} are expressed as
\begin{equation}
\begin{aligned}
-\mathcal{L} _{Y}^{int} =&\sum_{f=u,d,e} \frac{m_{f} }{v } \left ( \xi _{h}^{f}\bar{f}fh+\xi _{H}^{f}\bar{f}fH-2iI_{f}\xi _{f} \bar{f}\gamma _{5}fA \right )\\&+\frac{\sqrt{2}}{v } \left [V _{ud}\bar{u}\left ( m_{d} \xi _{d}P_{R}-m_{u} \xi _{u}P_{L}  \right )dH^{+}+m_{e}\xi _{e}\bar{\nu }P_{R}eH^{+}+h.c. \right]\changed{}{,}
\label{eq:fcouping}
\end{aligned}
\end{equation}
where $ \xi _{h}^{f}$ and $\xi _{H}^{f}$ are defined by
\begin{equation} 
\xi _{h}^{f}=s_{\beta - \alpha } +\xi _{f} c_{\beta - \alpha }, \quad
\xi _{H}^{f}=c_{\beta - \alpha } -\xi _{f} s_{\beta - \alpha },
\label{eq:xi}
\end{equation}
with $I_u=1/2$, and $I_{d,e}=-1/2$.
The values of $\xi _{u,d,e}$ in different types of 2HDM are shown in \autoref{tb:couping}.  Given the difference in the fermion couplings under four different types, the results on the constrained parameter space based on the charged Higgs searches vary a lot.


\section{Experimental constraints} \label{sec3}
 In this chapter, we summarize the current available direct search results of the charged Higgs from the LHC and the LEP, as well as flavor measurements on $B$ physics from the LHCb, BaBar and Belle.   

\paragraph{Conventional channels:} Conventional searches for the charged Higgs boson at the LHC  focused on the decay into fermions: $H^\pm \rightarrow \tau\nu$ ~\cite{ATLAS:2012tny,ATLAS:2012nhc,ATLAS:2014otc,ATLAS:2016avi,ATLAS:2018gfm,CMS:2012fgz,CMS-PAS-HIG-14-020,CMS-PAS-HIG-13-026,CMS:2015lsf,CMS:2019bfg,CMS:2016szv}, $cs$~\cite{ATLAS:2013uxj,CMS:2015yvc,CMS:2020osd,ATLAS:2024oqu}, $cb$~\cite{ATLAS:2023bzb,CMS:2018dzl} and $tb$~\cite{ATLAS:2015nkq,ATLAS:2018ntn,ATLAS:2021upq,CMS:2014pea,CMS:2015lsf,CMS:2020imj,CMS:2019rlz}, which are collected in \autoref{tb:ffdecay}.  Measurements of the non-resonant search of the SM process $ttbb$ have been performed at both ATLAS and CMS~\cite{ATLAS:2018fwl, CMS:2019eih, CMS:2020grm, CMS:2020utv, CMS:2023plc, CMS:2023xjh, Pfeffer:2024atk, ATLAS:2024aht}, along with   modeling studies~\cite{ATLAS:2022uiy, ATLAS:2022hmz, Ferencz:2023fso} since it gives irreducible background to the top pair associated Higgs boson production ($t\bar{t}H/A$) with $H/A\rightarrow bb$. We note that such continuum measurement can constrain the $tbH^\pm$ production with $H^\pm \rightarrow tb$ as well, especially when the width of the charged Higgs is large such that resonant search becomes ineffective. The current experimental constraints on the   $ttbb$ channel depends on the definitions of their fiducial phase space regions which usually involves some non-trivial cuts. Therefore, without detailed collider analyses it is difficult to reinterpret these constraints on the parameter space of a charged Higgs. However, we want to emphasize that such an analysis may be powerful, similar to the impact of continuum $tttt$ analysis on the neutral scalar searches~\cite{Kling:2020hmi}. 

\begin{table}[!ht]
\centering
\begin{tabular}{|c|c|c|c|c|c|c|}
\hline
\multirow{2}{*}{channel} & \multicolumn{3}{c|}{ATLAS} & \multicolumn{3}{c|}{CMS} \\
 \cline{2-7}                           
                         & 7 TeV   & 8 TeV   & 13 TeV   & 7 TeV   & 8 TeV  & 13 TeV  \\
\hline                         
$H^{\pm } \rightarrow\tau\nu$ 
&~\cite{ATLAS:2012tny,ATLAS:2012nhc} & ~\cite{ATLAS:2014otc} &~\cite{ATLAS:2016avi,ATLAS:2018gfm} & ~\cite{CMS:2012fgz}  &  ~\cite{CMS-PAS-HIG-14-020,CMS-PAS-HIG-13-026,CMS:2015lsf} & ~\cite{CMS:2019bfg,CMS:2016szv}  \\
\hline
$H^{\pm } \rightarrow cs$ & ~\cite{ATLAS:2013uxj} & - & ~\cite{ATLAS:2024oqu} & - & ~\cite{CMS:2015yvc}  & ~\cite{CMS:2020osd}   \\
\hline
$H^{\pm } \rightarrow cb$ & - & - & ~\cite{ATLAS:2023bzb} & - &   ~\cite{CMS:2018dzl} &   -    \\
\hline
$pp \rightarrow tbH^{\pm } \rightarrow ttbb$ &   -     &  ~\cite{ATLAS:2015nkq}      &  ~\cite{ATLAS:2018ntn,ATLAS:2021upq} &  -   &  ~\cite{CMS:2014pea,CMS:2015lsf} &  ~\cite{CMS:2020imj,CMS:2019rlz} \\
\hline
\end{tabular}
\caption{Charged Higgs searches with $H^{\pm} \rightarrow f f^\prime$ at the LHC. }
\label{tb:ffdecay}
\end{table}

\paragraph{Exotic decays channels:} The exotic decay channels $H^{\pm } \rightarrow W^{\pm(*)}\phi(\phi=A/H/h)$ open when the mass of the charged Higgs is greater than the mass of the neutral Higgs.   The current searches results at the LHC for $H^{\pm } \rightarrow AW^{\pm }/HW^{\pm }$ are very limited, as listed in ~\autoref{table:HpmtoAW}. For the light charged Higgs,  CMS results are only given for the cases $m_{H^{\pm }} = m_{A}+85$ GeV with a luminosity of 35.9 ${\rm fb^{-1}}$~\cite{CMS:2019idx}, while ATLAS only lists the results for $m_{H^{\pm }}$ = 120, 140, 160 GeV with a luminosity of 139 ${\rm fb^{-1}}$~\cite{ATLAS:2021xhq}.  Both collaborations focus on the final states of $W\mu\mu$. For the heavy charged Higgs, CMS also performed a search with an integrated luminosity of 138 ${\rm fb^{-1}}$ in the $H^{\pm } \rightarrow W^{\pm } H \rightarrow W\tau\tau$ channel assuming $m_{H} = 200$ GeV~\cite{CMS:2022jqc}.  

\begin{table}[!ht]
\centering
\begin{tabular}{|c|c|c|c|}
\hline
 \multirow{2}{*}{mass}&\multirow{2}{*}{channel}& {ATLAS} & {CMS} \\
  \cline{3-4}                           
                        & & 13TeV   & 13TeV  \\
\hline                         
$m_{H^{\pm}}< m_t$&~$H^{\pm } \rightarrow AW \rightarrow W\mu\mu$ 
 &~\cite{ATLAS:2021xhq}  &~\cite{CMS:2019idx}   \\
\hline
$m_{H^{\pm}}> m_t$&$H^{\pm } \rightarrow HW \rightarrow W\tau\tau$ 
 & -  & ~\cite{CMS:2022jqc}   \\
\hline
\end{tabular}
\caption{Charged Higgs searches with $H^{\pm} \rightarrow HW/AW$ at the LHC. }
\label{table:HpmtoAW}
\end{table}

\paragraph{LEP searches:} Searches for the charged Higgs bosons have been performed at the LEP, mainly using the $H^\pm$-pair production $e^{+ } e^{- }\rightarrow H^{+ } H^{- }$, since the cross section for $e^{+ } e^{- }\rightarrow H^{\pm  } W^{\mp  }$ is suppressed~\cite{Kanemura:1999tg}. In the 2HDM, the production cross section of charged Higgs boson pair is completely determined at the tree-level by the charged Higgs boson mass and known SM parameters~\cite{Djouadi:1992pu}. The limits on the charged Higgs boson mass are robust when all relevant charged Higgs boson decay channels are considered. Combining data from the four collaborations ALEPH~\cite{ALEPH:2002ftu}, DELPHI~\cite{DELPHI:1999yui}, L3~\cite{L3:2003jyb}, and OPAL~\cite{OPAL:1998dmf} on  $H^\pm \rightarrow \tau\nu$ and $H^\pm \rightarrow cs$ channels, a limit of $m_{H^{\pm } } \ge 80$ GeV is obtained under the assumption that  $H^{\pm }\rightarrow AW^{\pm}$ is absent.  The DELPHI~\cite{DELPHI:2003eid} and OPAL~\cite{OPAL:2008tdn} collaborations also searched for $H^{\pm }\rightarrow AW^{\pm}$. In the Type-I 2HDM, for $m_{A } = 12$ GeV, a limit of $m_{H^{\pm } } \ge 72.5$ GeV is obtained~\cite{ALEPH:2013htx}.  In this work we use LEP limits from the $\tau\nu$ final state and the $cs$ final state, and the OPAL limits from the $AW^{\pm }$ channel, with data obtained from \textbf{HiggsBounds}~\cite{Bechtle:2008jh,Bechtle:2011sb,Bechtle:2012lvg,Bechtle:2013wla,Bechtle:2015pma,Bechtle:2020pkv}.

\begin{table}[!ht]
\centering
\begin{tabular}{|c|c|c|}
\hline
Observable & Experimental result ~ & SM prediction ~ \\  
\hline
\cline{2-3}
$R_{\gamma}$ & $(3.22\pm 0.15)\times 10^{-3}$~\cite{Misiak:2017bgg} ~ & $(3.35\pm 0.16)\times 10^{-3}$~\cite{Misiak:2020vlo} ~  \\ 
\hline
\cline{2-3}
$BR(B\to\chi _{s} \gamma)$ & $(3.32\pm 0.15)\times 10^{-4}$~\cite{HFLAV:2019otj} ~ & $(3.40\pm 0.17)\times 10^{-4}$~\cite{Misiak:2020vlo} ~  \\ 
\hline
\cline{2-3}
$BR(B\to\tau\nu)$ & $(1.094\pm 0.208)\times10^{-4}$~\cite{Banerjee:2024znd} ~ & $(9.24 \pm 1.13)\times10^{-5}$~\cite{Atkinson:2021eox} ~  \\ 
\hline
\cline{2-3}
$BR(B_s\to\mu^+\mu^-)$& $(3.34 \pm  0.27)\times10^{-9}$~\cite{ParticleDataGroup:2024cfk} ~ &$(3.64\pm 0.12)\times10^{-9}$~\cite{Czaja:2024the} ~  \\ 
\hline
\cline{2-3}
$BR(B_d\to\mu^+\mu^-)$& $(0.6\pm 0.7)\times10^{-10}$~\cite{Chen:2024cqn} ~ &$(1.03\pm 0.05)\times10^{-10}$~\cite{Chen:2024cqn}  ~  \\ 
\hline
\cline{2-3}
$BR(D_s\to\tau\nu)$ & $(5.33\pm 0.12)\times10^{-2}$~\cite{Banerjee:2024znd} ~ & $(5.22\pm 0.04)\times10^{-2}$~\cite{Atkinson:2021eox} ~  \\ 
\hline
\cline{2-3}
$BR(D_s\to\mu\nu)$& $(5.36\pm 0.12)\times10^{-3}$~\cite{Banerjee:2024znd} ~ & $(5.31\pm 0.04)\times10^{-3}$~\cite{Atkinson:2021eox} ~  \\ 
\hline
\cline{2-3}
$R(D)$& $( 0.342\pm 0.026)$~\cite{Banerjee:2024znd} ~ & $(0.303\pm 0.006)$~\cite{Atkinson:2021eox} ~  \\ 
\hline
\cline{2-3}
\end{tabular}
\caption{Experimental results for   flavor physics observables and their corresponding SM predictions. }
\label{table:flavour}
\end{table}

\paragraph{Flavor constraints:} One of the  most stringent constraints on the 2HDM comes from the precision measurements of flavor
physics observables, especially the inclusive decay $B\to \chi _{s} \gamma$. At $E_{0} =1.6$ GeV, the experimental world average for $BR(B\to\chi _{s} \gamma)$ is $(3.32\pm 0.15)\times 10^{-4}$~\cite{HFLAV:2019otj}, and $R_{\gamma}^{exp }= B_{(s+d)\gamma } /B_{cl\bar{\nu }}=(3.22\pm 0.15)\times 10^{-3}$~\cite{Misiak:2017bgg}. The latest update of the SM values at the NNLO in QCD are~\cite{Misiak:2020vlo}, 
\begin{equation} \label{eq:flavor}
\begin{split}
BR(B\to\chi _{s} \gamma)^{SM } =(3.40\pm 0.17)\times 10^{-4}, ~~~  ~~~  R_{\gamma}^{SM } =(3.35\pm 0.16)\times 10^{-3}.
\end{split}
\end{equation}
We also investigate the constraints on the charged Higgs from other flavor physics observations. Both the experimental measurements and their corresponding SM predictions are collected in \autoref{table:flavour}.
Employing the method outlined in Ref.~\cite{Misiak:2020vlo}, we calculate the theoretical value of $R_{\gamma}$ and $BR(B\to\chi _{s} \gamma)$ in the 2HDM. In addition, we use the \textbf{SuperIso}~\cite{Mahmoudi:2009zz, Mahmoudi:2008tp, Mahmoudi:2007vz} to calculate the other flavor observables in the 2HDM.

\begin{figure}[htbp]
  \includegraphics[width=0.5\linewidth]{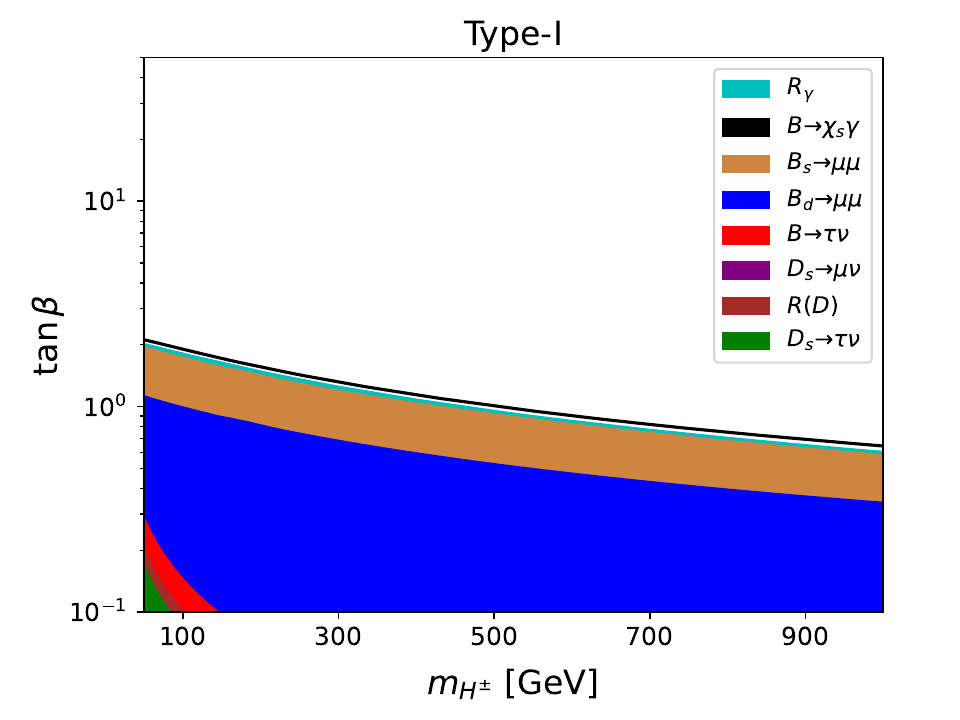}
  \includegraphics[width=0.5\linewidth]{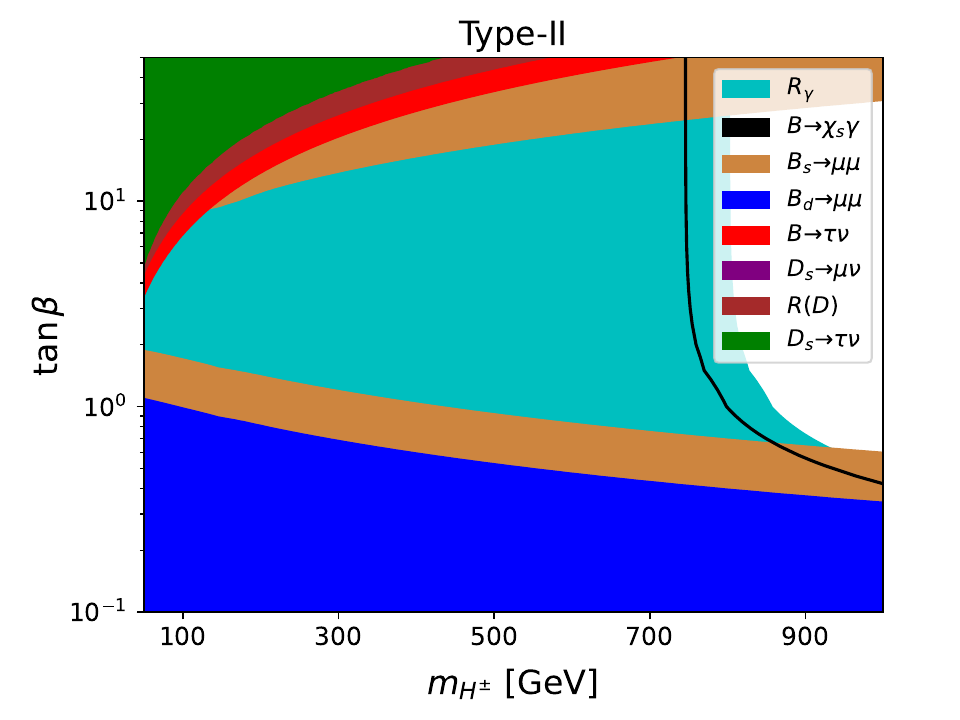}\\
  \includegraphics[width=0.5\linewidth]{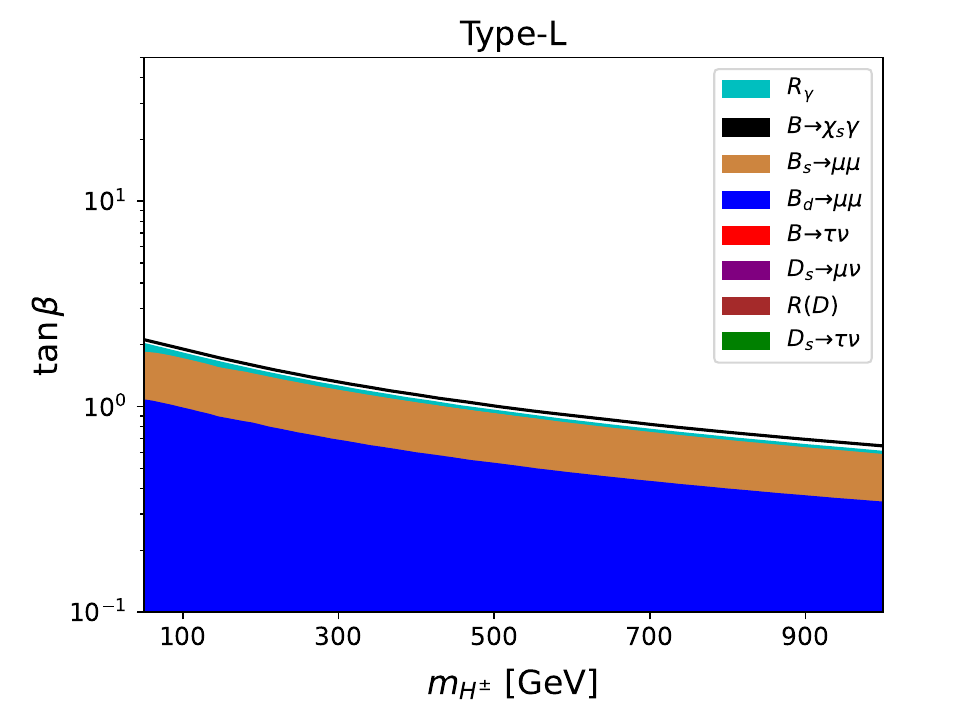}
  \includegraphics[width=0.5\linewidth]{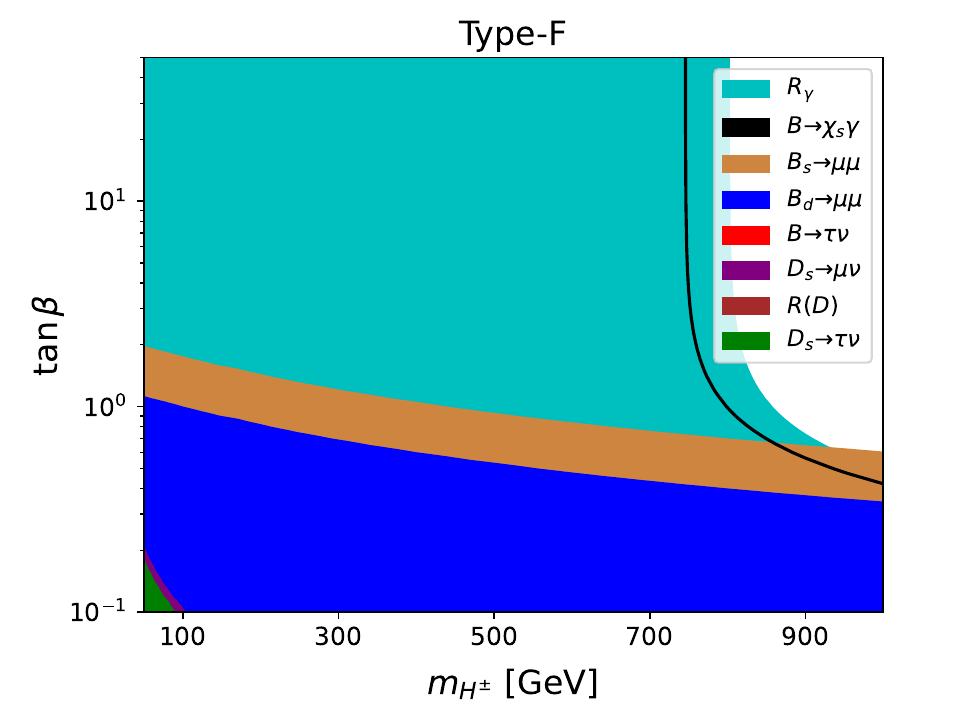}
  \caption{95\% C.L. contour plots of the allowed parameter space in the $m_{H^{\pm }}$ vs. $\tan \beta$ plane for Type-I 2HDM (upper left), Type-II 2HDM (upper right), Type-L 2HDM (lower left) and Type-F 2HDM (lower right) under flavor constraints.   }
 \label{fig:flavor}
 \end{figure}

 In Figure \ref{fig:flavor}, we present the flavor constraints on different types of 2HDM.  For the Type-I 2HDM (upper left panel), $ B\to \chi_{s} \gamma $ constraints are sensitive only to low $\tan \beta $ around 2 or less, given that all fermion couplings are proportional to $1/\tan\beta$.  For the Type-II 2HDM (upper right panel), the measurements of $R_{\gamma } = B_{(s+d)\gamma } /B_{cl\bar{\nu }}$ impose a very strong constraint on the Type-II 2HDM, leading to a lower limit on the mass of the charged Higgs particle to be around 800 GeV. For the Type-L (Type-F) 2HDM, the strongest constraints imposed by flavor physics on the charged Higgs are similar to those of Type-I (Type-II). Even though in the Type-II and Type-F 2HDM, charged Higgs mass less than about 800 GeV has been tightly constrained by the flavor measurements, in our collider analyses below, we consider all ranges of charged Higgs mass as an independent approach for the charged Higgs sector.

\paragraph{Precision constraints.} Electroweak precision measurements~\cite{ALEPH:2005ab} require the mass of the charged Higgs to be close to the mass of one of the neutral Higgses, $m_{H^{\pm }} \sim m_{H}$  or $m_{A}$~\cite{Kling:2016opi}. In our analyses below, we take the assumption of $m_{H^{\pm}}=m_H$ to satisfy the electroweak precision constraints.  The case where $m_{H^{\pm}}=m_A$ can be analyzed similarly. 


\section{Higgs searches} \label{sec4}
To interpret the experimental results, we take the experimental upper limits on the cross sections multiplied by the branching fractions $\sigma \times BR$ for various channels, as well as SM-like Higgs precision measurements and $\Gamma _{H^{\pm}}$ to directly constrain the 2HDM parameter space. We use the \textbf{prospino}~\cite{Beenakker:1996ed, Plehn:2002vy} package to calculate  the charged Higgs production cross sections in the 2HDM at the NLO level and the \textbf{2HDMC}~\cite{Eriksson:2009ws}  to calculate the Higgs decay branching fractions.

\subsection{Degenerate Higgs masses} \label{sec4.1}
\begin{figure}[htbp]
  \includegraphics[width=0.5\linewidth]{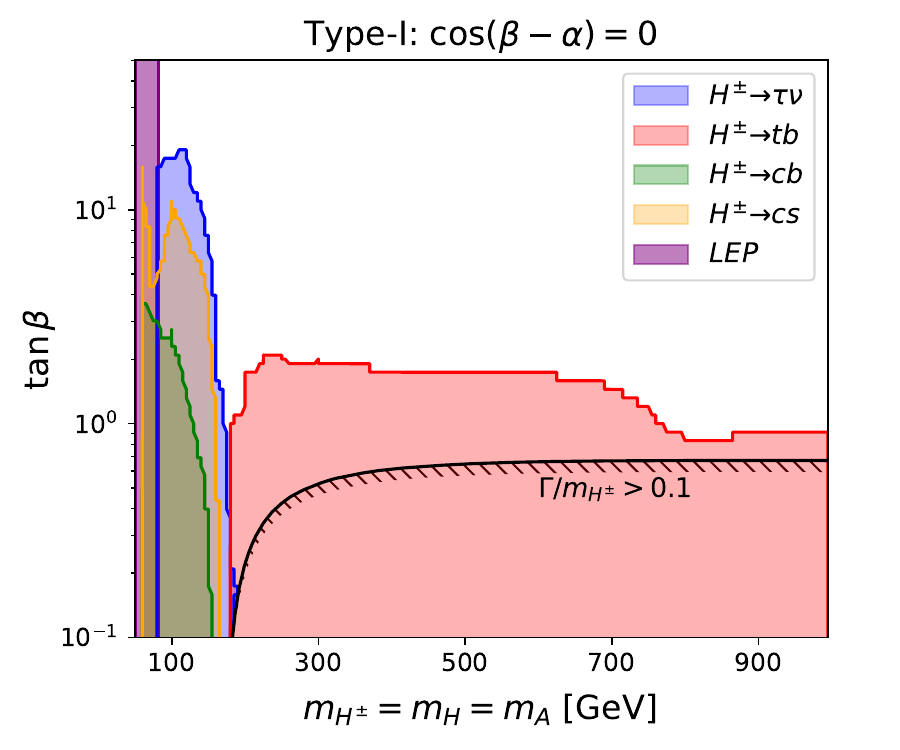}
  \includegraphics[width=0.5\linewidth]{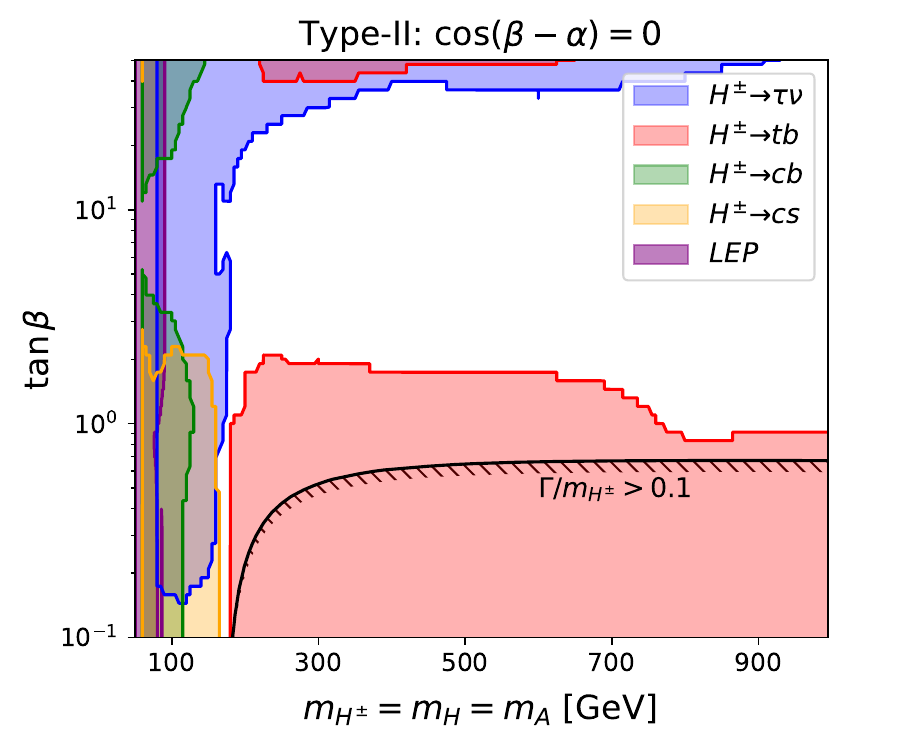}
  \includegraphics[width=0.5\linewidth]{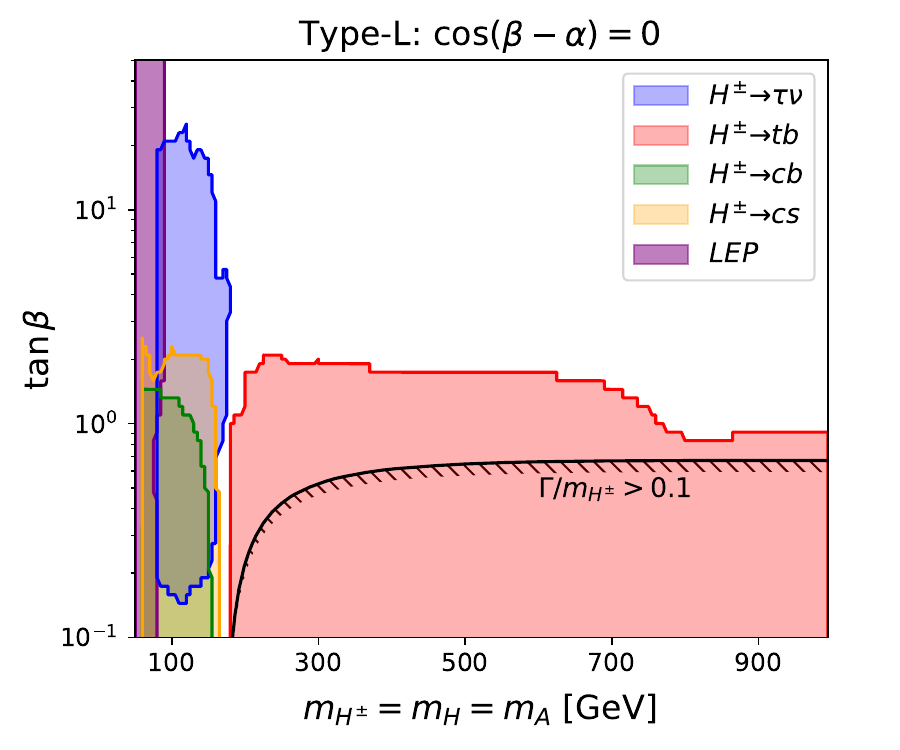}
  \includegraphics[width=0.5\linewidth]{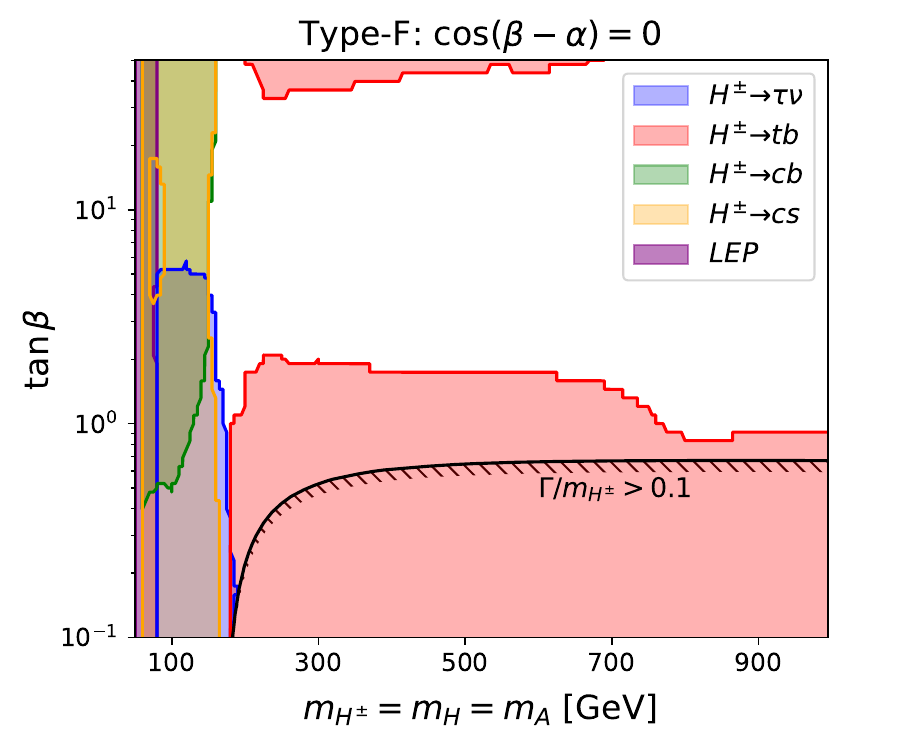}
  \caption{The 95$\%$ C.L. exclusion regions in the $m_{H^{\pm }}$ vs. $\tan \beta$ plane in the Type-I 2HDM (upper left), Type-II 2HDM (upper right), Type-L 2HDM (lower left) and Type-F 2HDM (lower right) with the degenerate mass spectrum of $m_{H^{\pm }} = m_{H} = m_{A}$ under alignment limit, taking into account i) the theoretical calcuations of the charged Higgs width $\Gamma /m_{H^{\pm } } > 0.1$ (hatch line); ii) the conventional search results at the LHC via $H^{\pm } \to \tau \nu$ (blue), $H^{\pm } \to cb$ (green), $H^{\pm } \to cs$ (orange), and $tbH^{\pm } \to ttbb$ (red);  iii) LEP limits on the charged Higgs (purple).   
  }
 \label{fig:de}
 \end{figure}

In Figure \ref{fig:de}, we show the constraints from direct searches for charged Higgs bosons at the LHC in the $m_{H^{\pm }}$ vs. $\tan \beta$ plane for four different types of the 2HDM under the alignment limit of $c_{\beta - \alpha }=0$. The $Z_{2}$ soft-break parameter $m_{12}^{2}$ is chosen to be $m_{12}^{2}=m_{H}^{2}s_{\beta }c_{\beta }$ to satisfy the theoretical constraints such as unitarity and vacuum stability conditions.  We assume a degenerate BSM Higgs spectrum $m_{H^{\pm }}= m_{H} = m_{A}$ such that electroweak precision constraints are automatically satisfied and the exotic decay channels do not open. The results are independent of $c_{\beta - \alpha }$ in the case of degenerate Higgs mass spectrum.

For the Type-I 2HDM (upper left panel), the couplings of charged Higgs and fermions are proportional to $1/\tan\beta$.  The charged Higgs productions, in low mass region of $m_{H^{\pm } } <  m_{t}$ via top decay $t \rightarrow b H^\pm$ and large mass region of $m_{H^{\pm } } >  m_{t}$ via $tbH^\pm$ associated production, are both suppressed at large $\tan\beta$.  Therefore, only relatively small $\tan\beta$ region is constrained by the direct searches.   In the low mass region, $H^\pm \rightarrow \tau\nu$ dominates, while the branching fraction of  $H^\pm \rightarrow cs$ is about a factor of two smaller.  The branching fraction of $H^\pm \rightarrow cb$ is suppressed by more than an order of magnitude.     The dominant decay channel of $H^\pm \rightarrow \tau\nu$  provides the most sensitive reach given the relatively clean final state, excluding $\tan\beta < 20$. Note that the low mass region is extended down to 60 GeV in the recent results from channel $H^\pm \rightarrow cb$~\cite{ATLAS:2023bzb} and $H^\pm \rightarrow cs$~\cite{ATLAS:2024oqu}. 
In the high mass region, $H^\pm$ dominantly decays to $tb$, suppressing the decay branching fractions of other decay channels by at least four orders of magnitude.  Therefore, $H^\pm \rightarrow tb$ provides the best limit of $\tan\beta\lesssim 1 - 2$ for $m_{H^\pm}<1$ TeV.   The slightly weaker constraint on the $tbH^{\pm } \to ttbb$ channel around $m_{H^{\pm } } = 800$ GeV is  due to upward fluctuation of the experimental data at that particular point~\cite{ATLAS:2021upq}.
We also show the region corresponding to  $\Gamma /m_{H^{\pm } } > 0.1$ (hatch line), in which the resonant search results are not applicable given the large charged Higgs decay width.   
Furthermore, LEP~\cite{ALEPH:2002ftu,DELPHI:1999yui,L3:2003jyb,OPAL:1998dmf,DELPHI:2003eid,OPAL:2008tdn} excludes a charged Higgs with mass up to 83 GeV via $e^{+}e^{-}\rightarrow H^{+}H^{-}$ production channel.

For the Type-II 2HDM (upper right panel), the charged Higgs couplings are proportional to $\tan\beta$ for the down quark-type Yukawa coupling, and $1/\tan\beta$ for the up quark-type Yukawa. The productions of charged Higgs via $t\to H^{\pm }b$ for small $m_{H^\pm}$ and $tbH^\pm$ for large $m_{H^\pm}$ are enhanced for both small and large $\tan\beta$.  For the light $H^\pm$,  $H^\pm \rightarrow \tau\nu$ dominates at almost 100\% branching fraction, which excludes $m_{H^\pm}<m_t$ at almost all values of $\tan\beta > 0.2$. The branching fraction of $H^\pm \rightarrow cb$ is larger than that of $cs$, which provides a better reach at small $\tan\beta$, and also provides a  certain exclusion at large $\tan\beta$ with the enhanced production rate from $t\to H^{\pm }b$. In the heavy $H^\pm$ case, while the small $\tan\beta$ exclusion is similar to that of the Type-I case via $H^\pm \rightarrow tb$,  $H^{\pm}\to \tau\nu$ provides a better reach at large $\tan\beta$ with the branching fraction of such channel being about 10\%. 
For $\tan\beta>30\ (40)$, the mass of charged Higgs $m_{H^\pm}$ has been excluded up to 315 (805) GeV. $H^\pm \rightarrow tb$ channel can also exclude a portion of the parameter space at very large values of $\tan\beta \sim 50$.

For the Type-L 2HDM (lower left panel), the couplings between the charged Higgs and quarks are the same as for the Type-I 2HDM, inversely proportional to the $\tan \beta$ value. Consequently, the production cross-section of the charged Higgs is suppressed at large values of $\tan \beta$ as in the Type-I 2HDM case. The coupling of the $H^{\pm }$ to $\tau \nu$, however, is proportional to $\tan \beta$.  Therefore, the decays of $H^{\pm } \to cb$ and $H^{\pm } \to cs$ are significantly suppressed with respect to the Type-I 2HDM when $\tan \beta \gtrsim 1$. The excluded region via $H^{\pm } \to \tau \nu$ is similar to the Type-II at small $\tan \beta$. For larger $\tan \beta$, there is no excluded region through the $\tau\nu$ channel since the production cross-section of the charged Higgs is heavily suppressed.

For the Type-F 2HDM (lower right panel), the couplings between the charged Higgs and quarks are the same as in the Type-II model, while the leptonic couplings are proportional to $1/\tan\beta$. Therefore, the production cross section of the charged Higgs is  enhanced at both small and large $\tan\beta$.  In the low mass region, the decay branching fraction of $H^{\pm } \to cs$ is around 0.2 $-$ 0.4 for all values of $\tan\beta$, which provides limits across all the values of $\tan\beta$. The decay mode of $H^{\pm }$ to $cb$ is significantly suppressed at small values of $\tan \beta$, while $H^{\pm } \to \tau \nu$ decay dominates.  However, when $\tan \beta$ exceeds 5, the decay channel $H^{\pm } \to cb$ becomes the dominant one, while the $H^{\pm } \to \tau \nu$ channel is significantly suppressed. Therefore, $H^\pm \rightarrow \tau\nu$ excludes light $H^\pm$ up to $\tan\beta \sim 5$ while $H^\pm \rightarrow cb$ excludes light $H^\pm$ for $\tan\beta > 0.4$.  For the heavy $H^\pm$,  the exclusion region is similar to that of the Type-II case via $H^\pm \rightarrow tb$, while $H^\pm \rightarrow \tau\nu$ exclusion disappears at large $\tan\beta$.

\subsection{Non-degenerate case} 
\label{sec4.2}
Once there are mass splittings between the charged Higgs and neutral Higgs masses, additional exotic channels such as $H^{\pm }\rightarrow AW/HW$ open and rapidly dominate the decay. 
Therefore, the reaches of the conventional searches shown earlier are  reduced. At the same time, the exotic decay channels $H^{\pm } \rightarrow AW/HW$ provide alternative possibilities for charged Higgs searches at the LHC. In our analysis below, we  assume that $m_{H^{\pm } }=m_H$ to satisfy the electroweak precision constraints. $H^{\pm } \rightarrow HW$ is therefore kinematically closed and our results are  independent of $c_{\beta -\alpha }$.  Once we relax the degenerate mass relation of $m_{H^{\pm } }=m_H$, $H^{\pm } \rightarrow HW$ could be kinematically accessible as well, which will introduce $c_{\beta -\alpha }$ dependence.  Similar analyses can be performed under the assumption of $m_{H^{\pm } }=m_A$ or more general spectrum of $m_{H^\pm}$, $m_H$ and $m_A$, as long as electroweak precision constraints are satisfied. Our analyses below under the assumption of $m_{H^{\pm } }=m_H$ nevertheless show the   general features of the complementarity between such exotic channels and the conventional charged Higgs search channels.

\subsubsection{$\tan\beta$ vs. $m_{H^{\pm}}=m_{H}$} 
\label{sec4.2.1}
\begin{figure}[htbp]
  \includegraphics[width=0.5\linewidth]{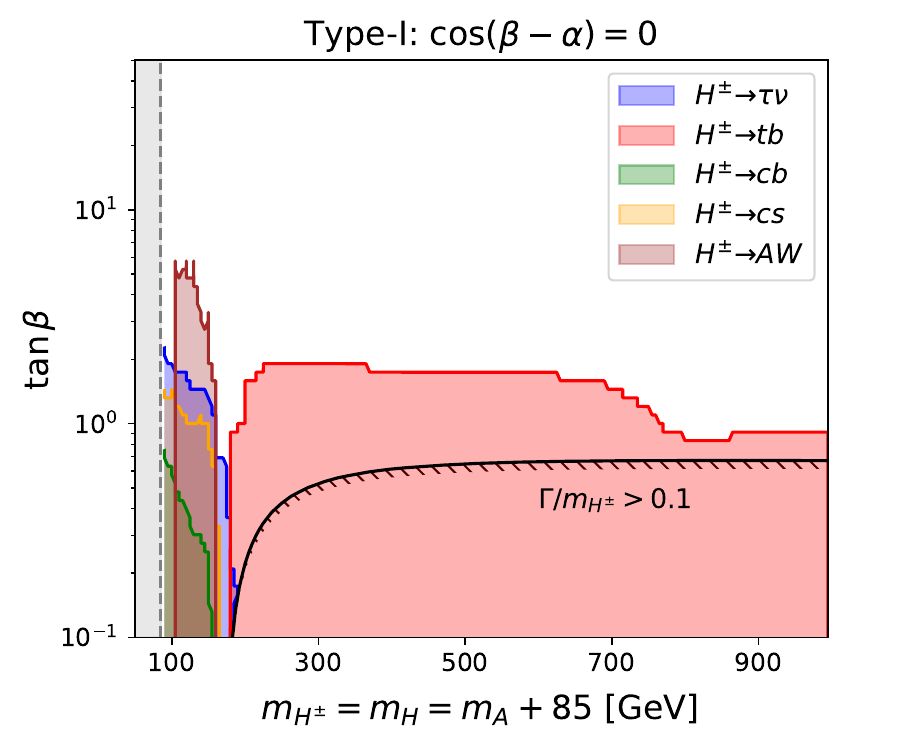}
  \includegraphics[width=0.5\linewidth]{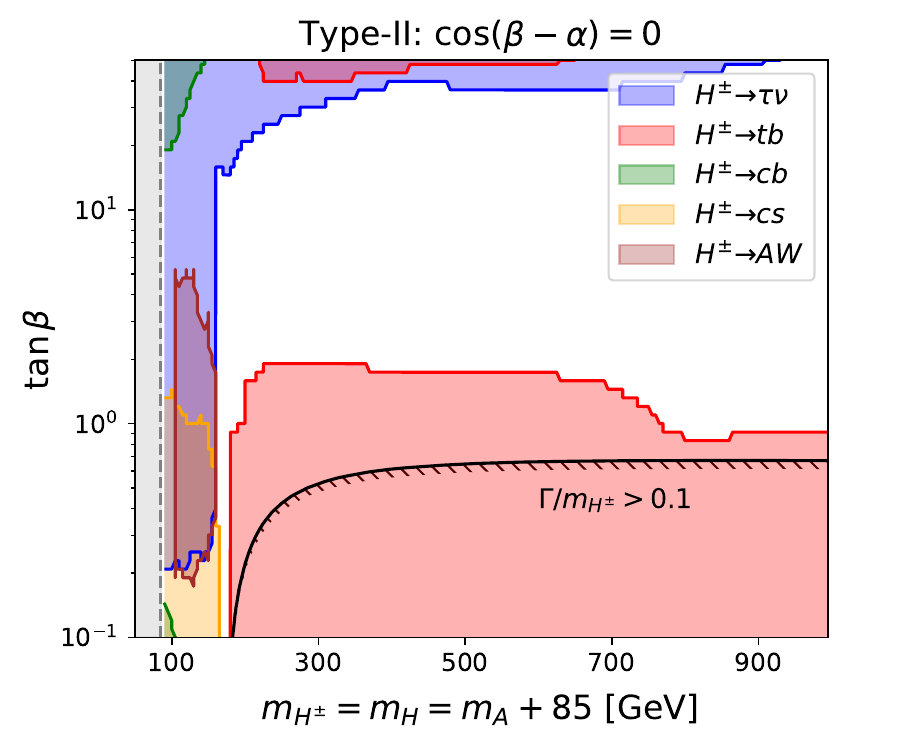}
  \includegraphics[width=0.5\linewidth]{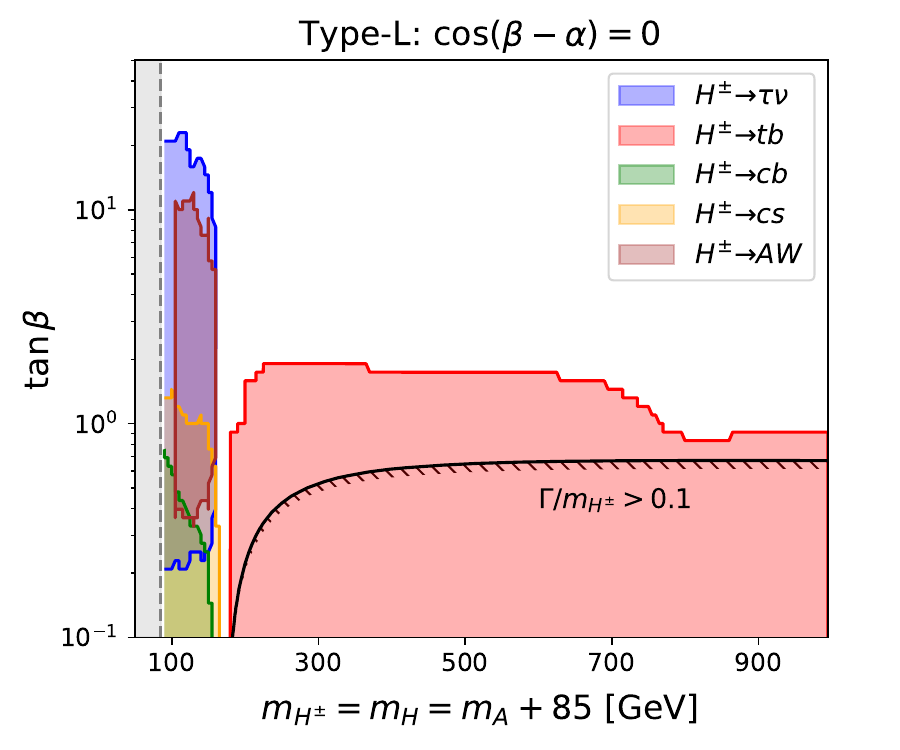}
  \includegraphics[width=0.5\linewidth]{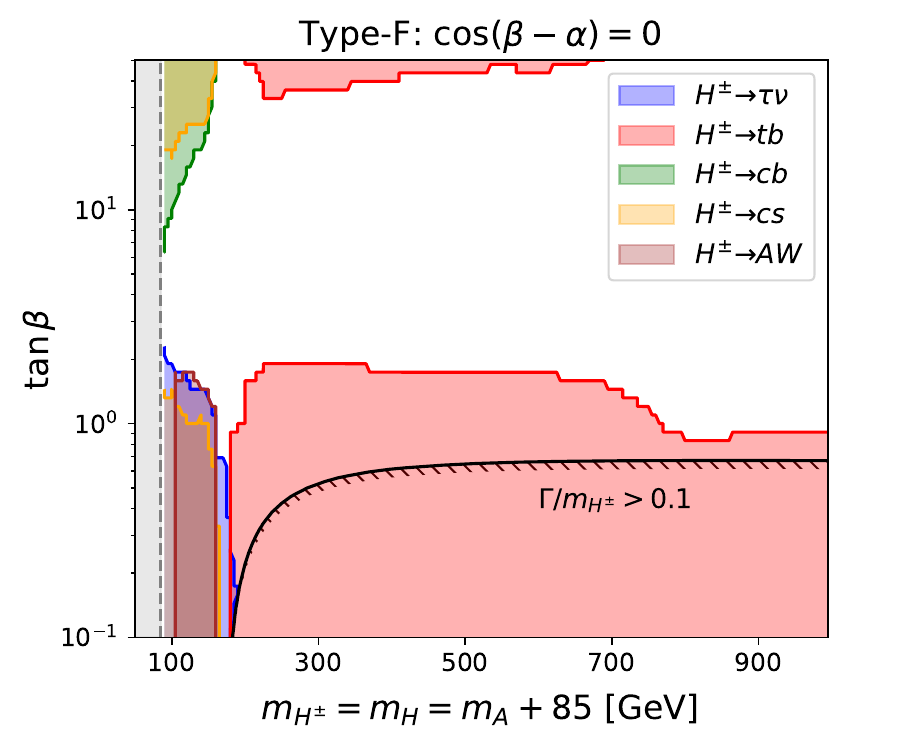}
  \caption{The 95$\%$ C.L. exclusion regions in the $m_{H^{\pm }}$ vs. $\tan \beta$ plane of the Type-I 2HDM (upper left), Type-II 2HDM (upper right), Type-L 2HDM (lower left) and Type-F 2HDM (lower right) with $\Delta m_A=m_{H^{\pm }}-m_{A}=85$ GeV. The constraints from $H^\pm \rightarrow AW$ is indicated in brown. Gray area indicates the nonphysical regions with negative values of $m_A$. The rest of color coding is the same as Figure~\ref{fig:de}.  }
 \label{fig:A85}
 \end{figure}
 
In Figure \ref{fig:A85}, we show the current collider limits~\cite{CMS:2019idx}  in the 2HDM on the $m_{H^{\pm }}$ vs. $\tan \beta$ plane, assuming that $\Delta m_A=m_{H^{\pm }}-m_{A}=85$ GeV in which the charged Higgs mass is just above the kinematic threshold of the decay mode $H^\pm \rightarrow AW$ (brown). The experimental results are only available for $m_{H^\pm}<m_t$. Gray area indicates the nonphysical regions with negative values of $m_A$.  

For the Type-I 2HDM (upper left panel), $H^\pm \to AW$ provides the strongest constraint for the light $H^\pm$ case with an upper limit of $\tan\beta$ around 6.  The reaches of $\tau\nu$, $cs$ and $cb$ channels are suppressed, comparing to those in the degenerate case.  Once $H^\pm \to tb$ opens up, it still dominates at small $\tan\beta$, with a collider reach similar to that in the degenerate case.

For the Type-II and Type-L 2HDM (upper right and lower left panels), with a mass difference of 85 GeV between $m_{H^{\pm }}$ and $m_{A}$, $H^\pm \rightarrow \tau\nu$ dominates over $AW$ channel for $\tan\beta \gtrsim 5$.  Therefore, the limit plots are very similar to those in the degenerate case, with only the addition of $H^\pm \rightarrow AW$ reach at light $H^\pm$, and the suppression of the reaches of $H^\pm \rightarrow cs, cb$. Note that the reach of the $AW$ channel only extends to $\tan\beta \sim 0.2 (0.4)$ for the Type-II (-L) 2HDM.  This is due to the suppression of decay channel $A \rightarrow \mu\mu$ in the Type-II (-L) 2HDM at small $\tan\beta$ for $A(\rightarrow\mu\mu)W$ channel.

For the Type-F 2HDM (lower right panel), the reaches from $H^\pm \rightarrow \tau\nu, cs$ and $cb$ at small $m_{H^\pm}$ are suppressed comparing to those in the degenerate case due to the opening of $H^\pm \rightarrow AW$. The reach of $H^\pm \rightarrow AW$, however, is reduced at $\tan\beta \gtrsim 2$ given the suppression of $A \rightarrow \mu\mu$.  For the heavy $H^{\pm}$, the exclusion region of the $tb$ channel is similar to that of the degenerate case.

\begin{figure}[htbp]
   \includegraphics[width=0.5\linewidth]{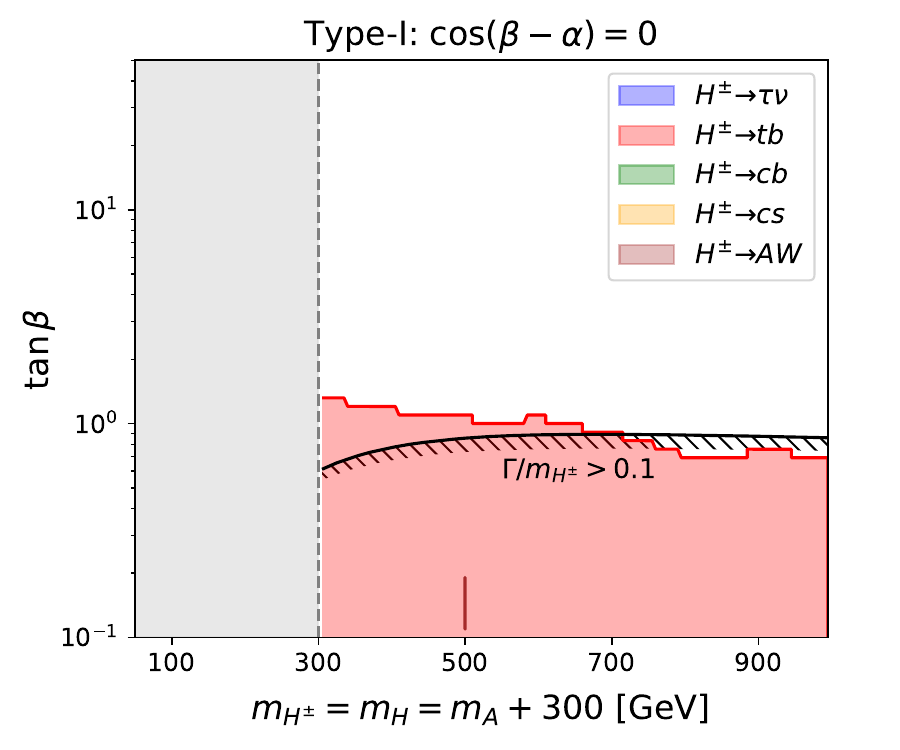}
   \includegraphics[width=0.5\linewidth]{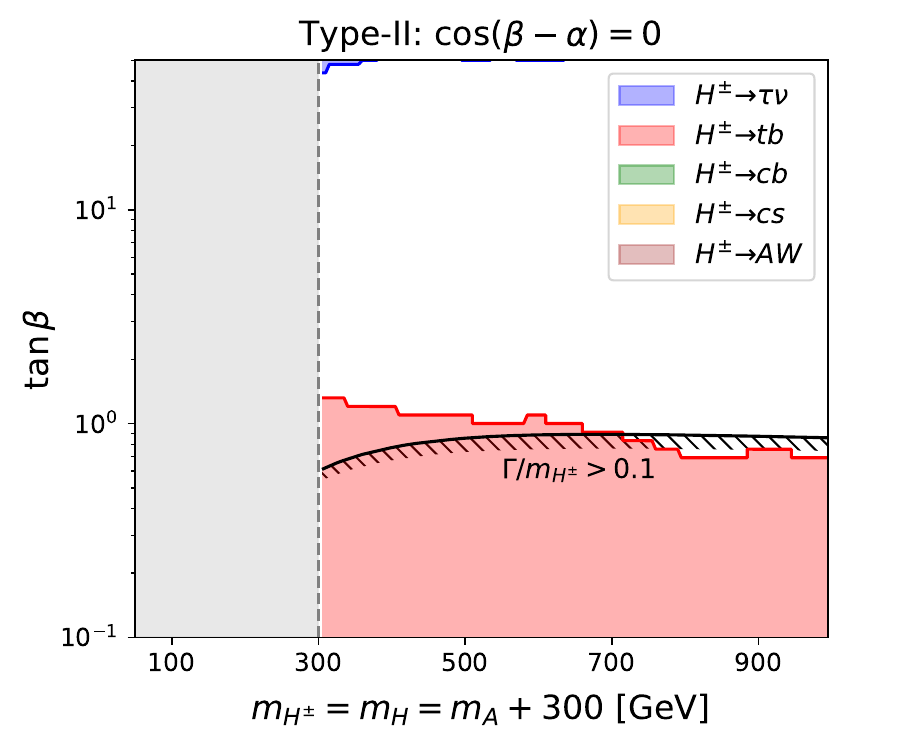}
   \includegraphics[width=0.5\linewidth]{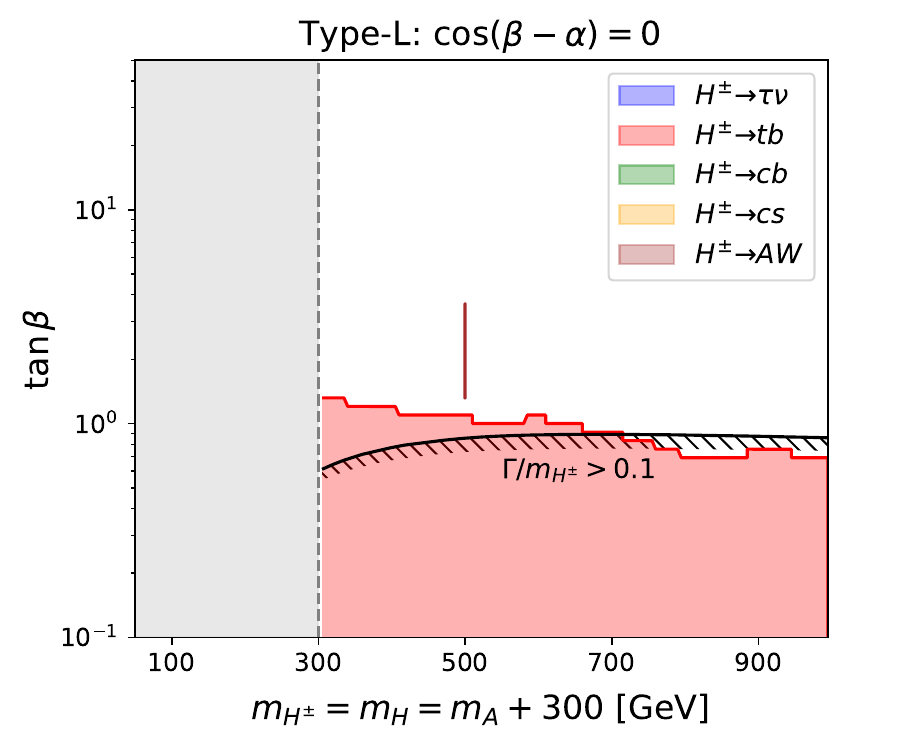}
   \includegraphics[width=0.5\linewidth]{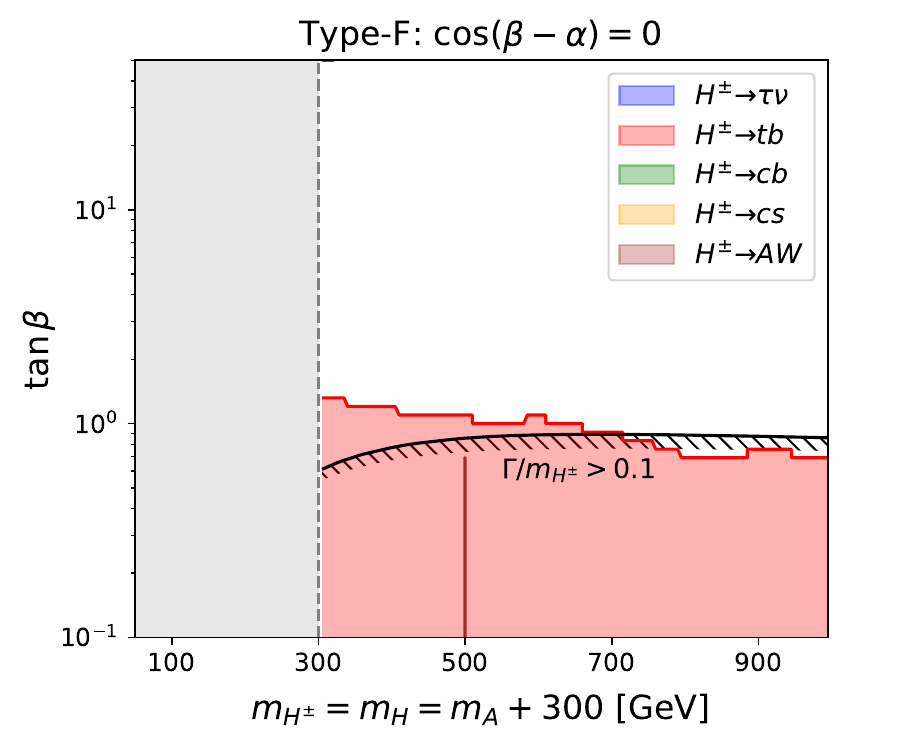}
   \caption{The 95$\%$ C.L. exclusion regions in the $m_{H^{\pm }}$ vs. $\tan \beta$ plane of the Type-I 2HDM (upper left), Type-II 2HDM (upper right), Type-L 2HDM (lower left) and Type-F 2HDM (lower right) with $\Delta m_A=m_{H^{\pm }}-m_{A}=300$ GeV. Color coding is the same as Figure~\ref{fig:A85}.}
 \label{fig:MHpm_tb_DeltamA300}
\end{figure}

In Figure \ref{fig:MHpm_tb_DeltamA300}, we show the current collider limits in the $m_{H^{\pm }} = m_{H}$ vs. $\tan \beta$ plane of four different types of 2HDM with $\Delta m_A=m_{H^{\pm }}-m_{A}=300$ GeV.  $m_{H^\pm}<300$ GeV is nonphysical for such $\Delta m_A$. Note that the experimental limits of $H^\pm \rightarrow A W$ channel  for $\Delta m_A \ne 85$ GeV are only given for a fixed value of $m_{A}=200$ GeV~\cite{CMS:2022jqc} for $m_{H^\pm}>m_t$.   Once $\Delta m_A \gtrsim 100$ GeV, $H^{\pm }\rightarrow AW$ quickly dominates over the fermionic decay channels. Therefore, the reach of $H^\pm \rightarrow tb$ channel at large $m_{H^\pm}$ is suppressed comparing to that in the degenerate case, as shown by the red regions in all panels. 

The best reach of $H^{\pm }\rightarrow AW$ is obtained in the Type-L 2HDM for $\tan\beta$ between 1 to 5, given the enhanced decay branching fraction of $A \rightarrow \tau\tau$.  For the Type-I and Type-F, the reach via $H^{\pm }\rightarrow AW$ is less comparing to that via $H^\pm \rightarrow tb$. There is no limit obtained from $H^{\pm }\rightarrow AW$ in the Type-II 2HDM due to the suppression of $A \to \tau\tau$ channel at small $\tan\beta$.  For larger mass splitting of $\Delta m_A$, the reach of $H^{\pm} \rightarrow tb$ is further suppressed, while the reach of $H^{\pm }\rightarrow AW$ slightly increases.  However, the feature remains similar to the case of $\Delta m_A=300$ GeV.

\subsubsection{$m_{H^{\pm }}$ vs. $m_A$} 
 \begin{figure}[htbp]
  \includegraphics[width=0.5\linewidth]{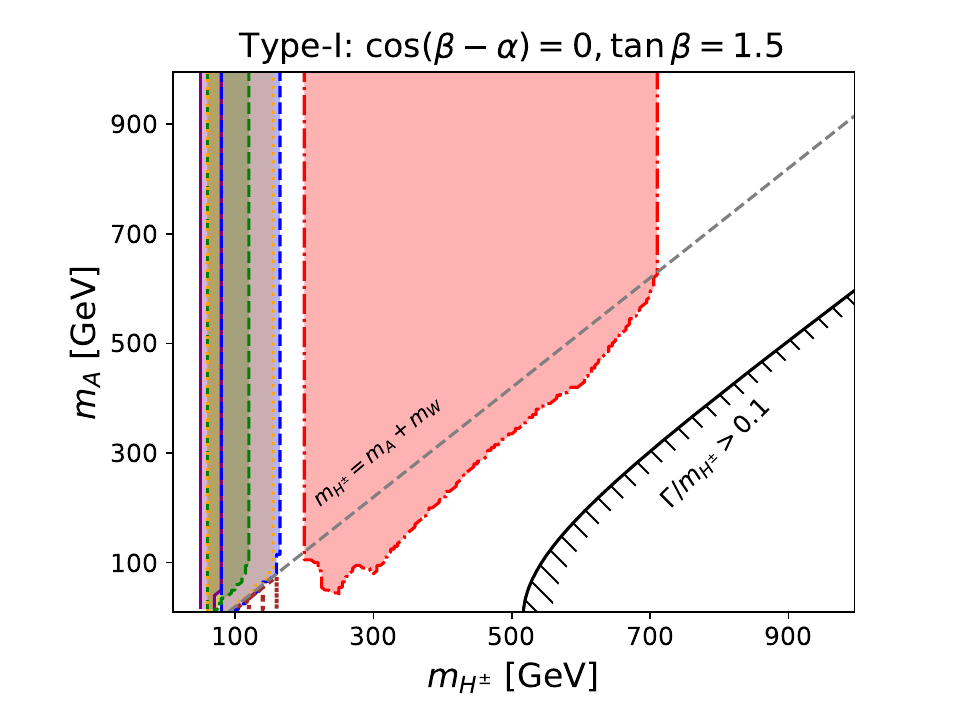}
  \includegraphics[width=0.5\linewidth]{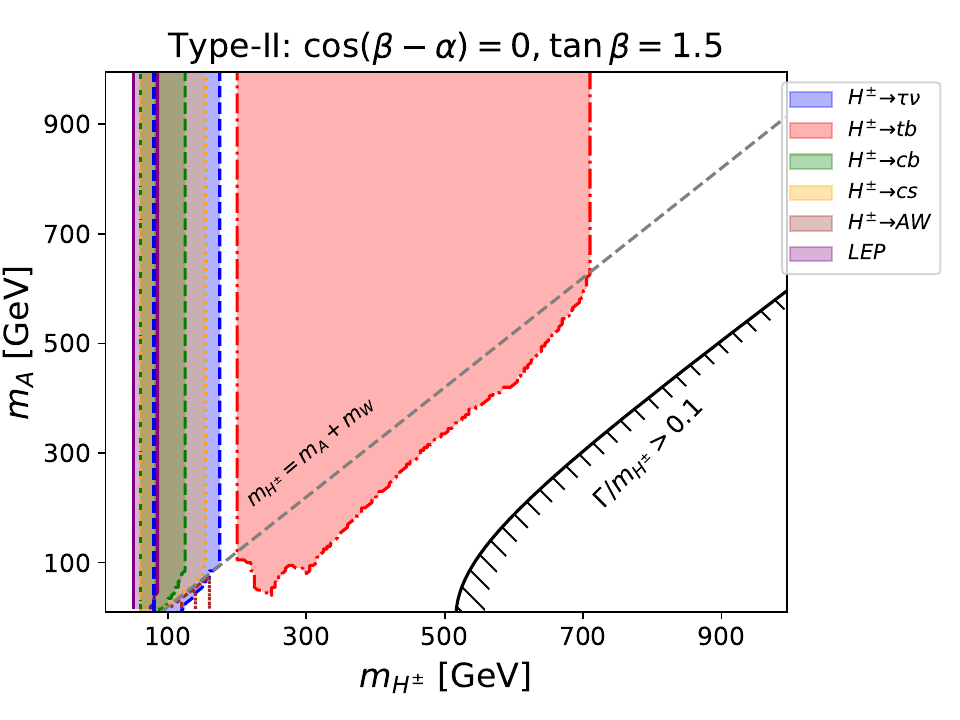}
  \includegraphics[width=0.5\linewidth]{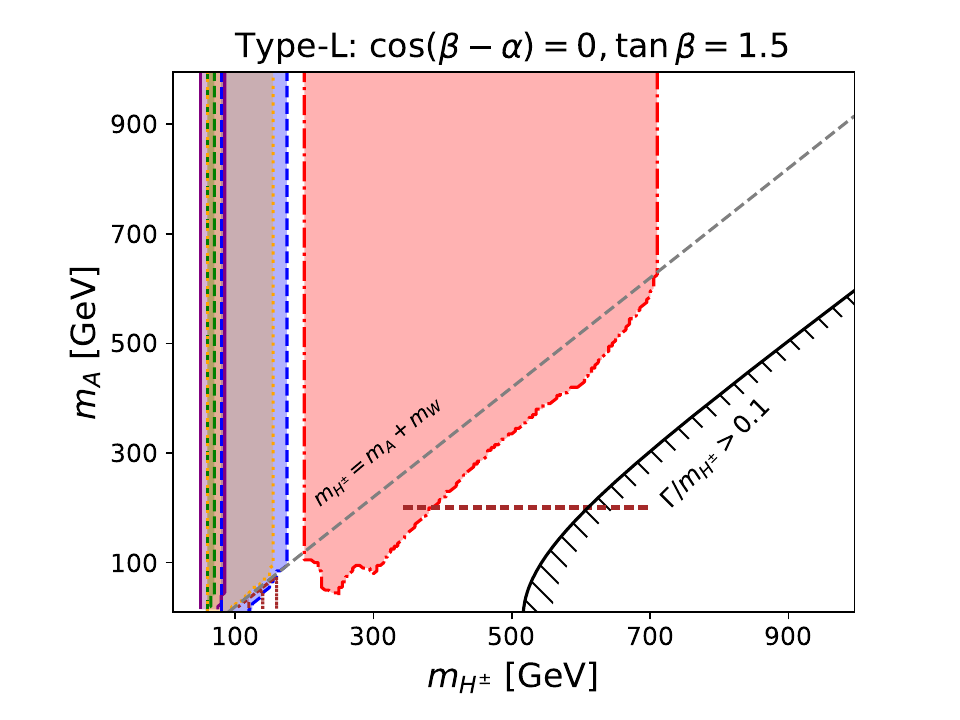}
  \includegraphics[width=0.5\linewidth]{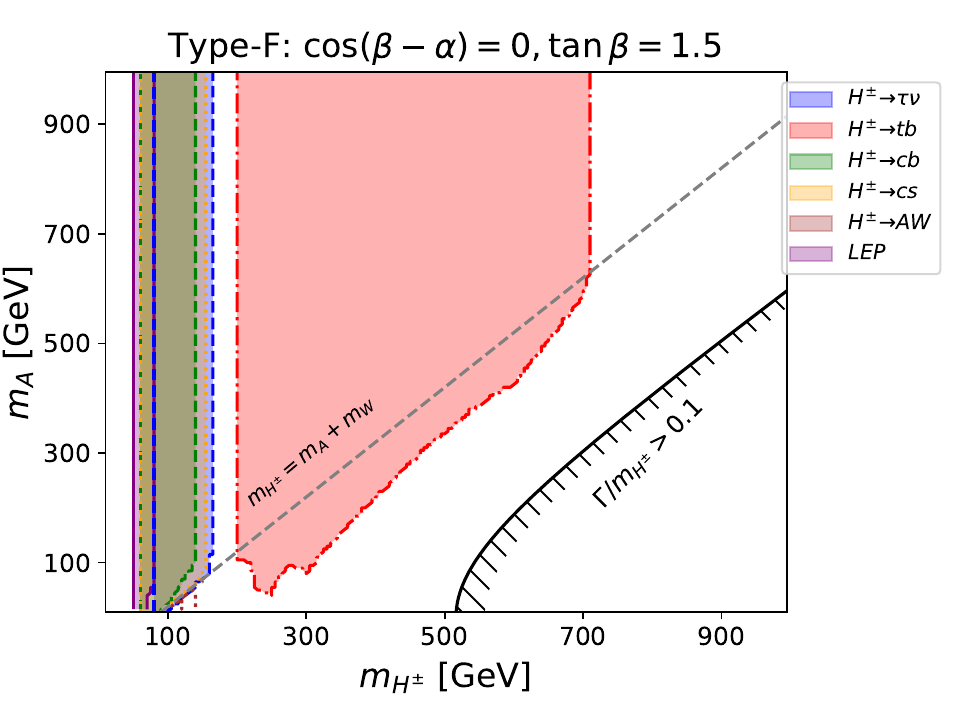}
  \includegraphics[width=0.5\linewidth]{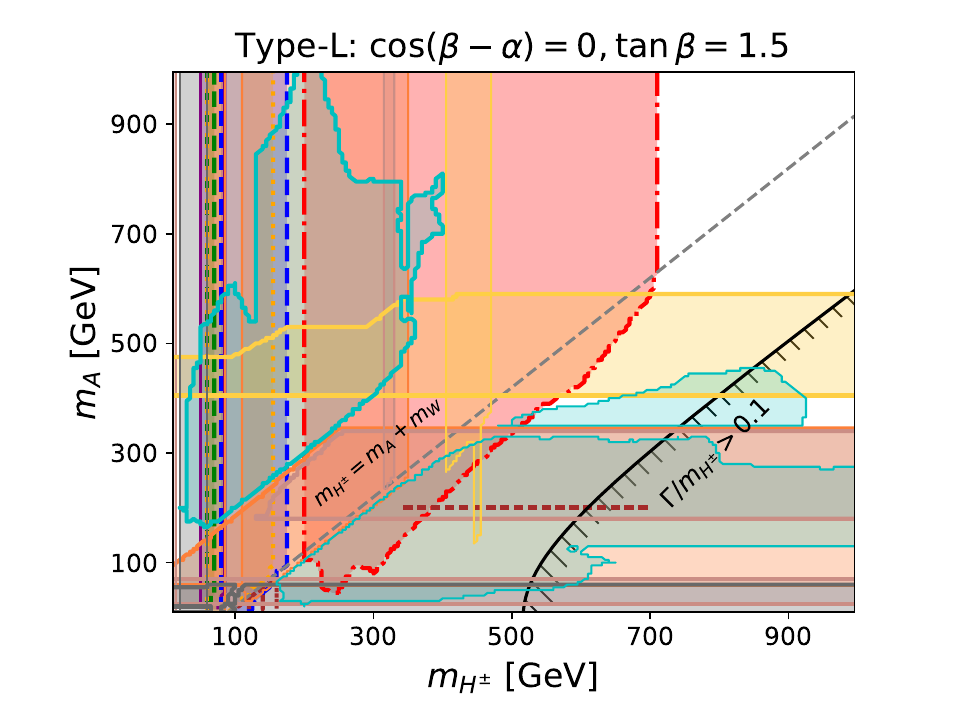}
  \includegraphics[width=0.5\linewidth]{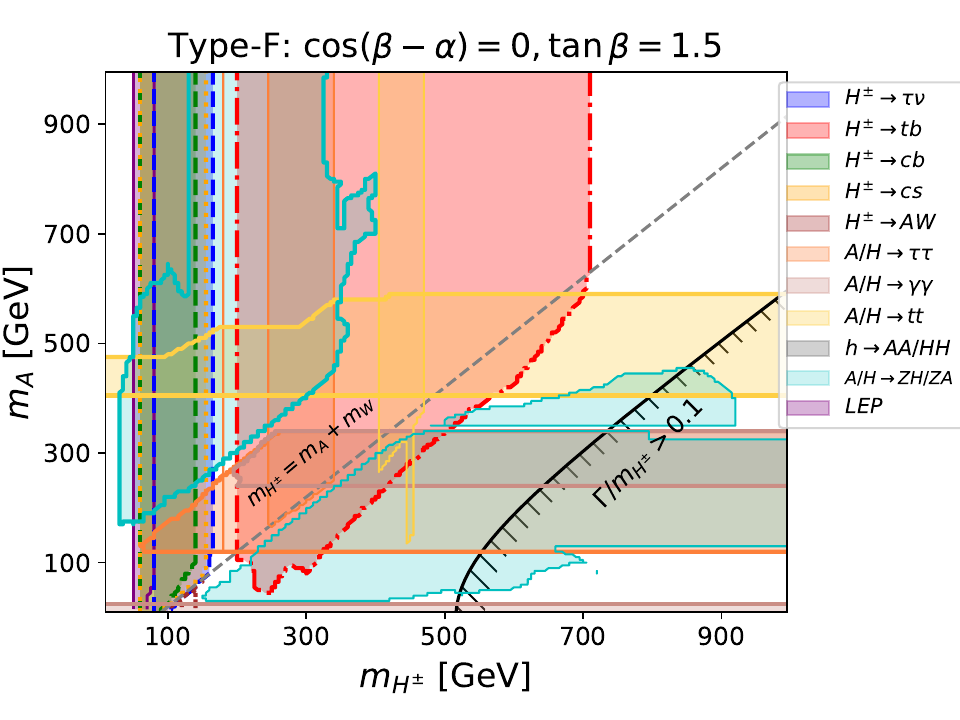}
  \caption{The 95$\%$ C.L. exclusion regions in the $m_{A}$ vs.$m_{H^{\pm }}$ plane for the Type-I 2HDM (upper left), Type-II 2HDM (upper right), Type-L 2HDM (middle and lower left) and Type-F 2HDM (middle and lower right) for $\tan \beta=1.5$. In the top four panels, we only show the constraints from $H^\pm$ searches. 
  In the bottom two panels, direct search constraints on the neutral scalar $A$ are included as well. Under the assumption of $m_{H^\pm}=m_H$, we also show regions excluded by the direct searches of $H$.  }
 \label{fig:tb1.5} 
 \end{figure} 

In Figure \ref{fig:tb1.5}, we show the LHC collider search limits in the $m_{H^{\pm }}$ vs. $m_A$ plane for $\tan \beta=1.5$.  In the top four panels we only take into account the charged Higgs searches.  Other than the region around $m_{H^\pm} \sim m_t$, which has a narrower gap in the Type-II and Type-L 2HDM due to the coverage from $H^\pm \rightarrow \tau\nu$, the exclusion regions of the conventional channels in all the four types of 2HDM are very similar.  The fermionic decays of $H^\pm$ almost cover all regions of $m_{H^\pm}\lesssim 715$ GeV for $m_{H^\pm}<m_A+m_W$,  except for $m_{H^\pm}\sim m_t$.  

The gray diagonal line indicates $m_{H^\pm}=m_A+m_W$,   below which the on-shell decay of $H^{\pm }\to AW$ is kinematically open. The $H^\pm tb$ associated production with     $H^\pm\rightarrow tb$   extends the reach beyond $\Delta m_A=m_W$.  Once $H^{\pm }\rightarrow AW$ is kinematically accessible, it provides the largest decay branching ratio of the charged Higgs, in particular for heavy $H^\pm$. 
The horizontal brown dashed line indicates $m_A=200$ GeV, which is the benchmark value of $m_A$ that is chosen for the $H^{\pm }\to AW$ search in the heavy $H^\pm$ case.    For the Type-L,  $H^{\pm }\rightarrow AW$ provides the best charged Higgs search limit. For the Type-I, -II, and -F, the reach of $H^{\pm }\to AW$ is severely suppressed due to the small branching ratio of $A\rightarrow\tau\tau$ at low $\tan\beta$.   

The bottom two panels show the exclusion reaches in the Type-L and -F including the collider search limits on the neutral scalar $A$ as well~\cite{Kling:2020hmi}, which are indicated by the horizontal exclusion bands. $A\rightarrow\tau\tau$~\cite{CMS:2018rmh,CMS:2019hvr,ATLAS:2020zms,CMS:2022goy} and $\gamma\gamma$~\cite{CMS:2024nht,ATLAS:2022abz,ATLAS:2018xad,ATLAS:2017ayi,ATLAS:2014jdv,CMS:2018cyk,CMS:2018dqv} dominate the reach below $2m_t$, while $A \rightarrow tt$~\cite{ATLAS:2024vxm,CMS:2019pzc} provides the best reach for 400 GeV $<m_A<580$ GeV. For the very small value of $m_A$, we include the limits from $h \rightarrow AA$~\cite{ATLAS:2021hbr,CMS:2020ffa,CMS:2021pcy,CMS:2022xxa,CMS:2022fyt,CMS:2024zfv,CMS:2018jid,ATLAS:2018coo,CMS:2018qvj,ATLAS:2015unc,CMS:2019spf,CMS:2018nsh,ATLAS:2018emt,CMS:2018zvv,ATLAS:2018pvw} as well. Note that if we impose the relation of $m_{H^\pm}=m_H$, additional constraints from $H$ searches (indicated by vertical bands), as well as $A\rightarrow ZH$~\cite{CMS:2016xnc,ATLAS:2018oht,CMS:2019ogx,ATLAS:2020gxx,ATLAS:2023szc,ATLAS:2022fpx}  and $H \rightarrow AZ$~\cite{CMS:2016xnc,ATLAS:2018oht,CMS:2019ogx,ATLAS:2020gxx,ATLAS:2023szc} also show up, which are indicated by the vertical and horizontal cyan region. $A/H \to HZ/AZ$ searches exclude regions of $m_{H/A}$ between 30 to 450 GeV.   Exclusions for the Type-I (II) are similar to those of the Type-F (L) from these additional $A$ and $H$ search modes.

\begin{figure}[htbp]
  \includegraphics[width=0.5\linewidth]{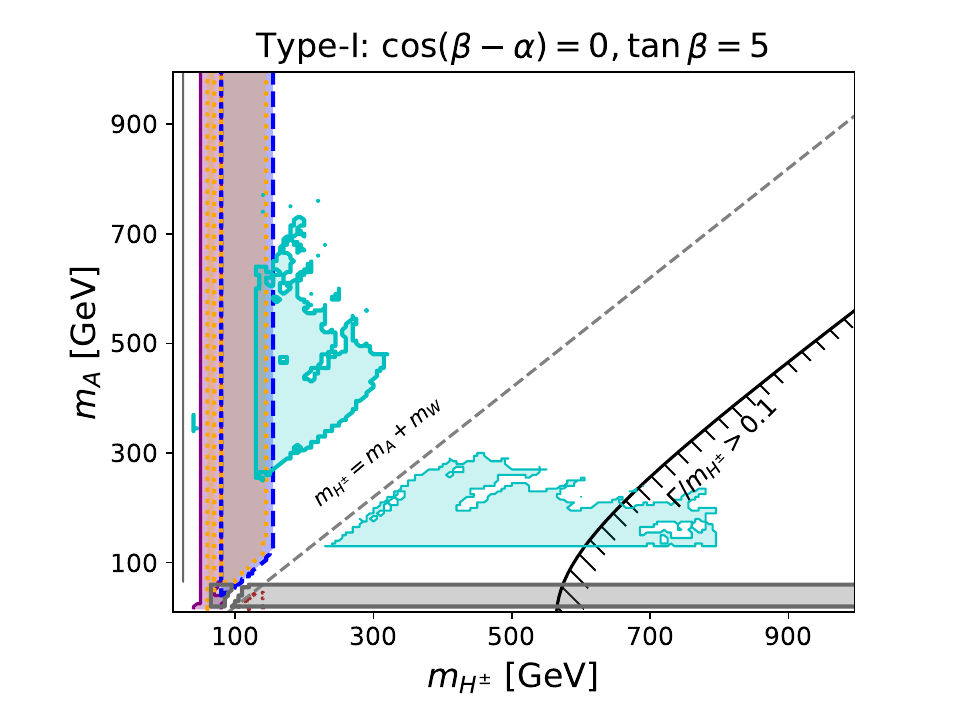}
  \includegraphics[width=0.5\linewidth]{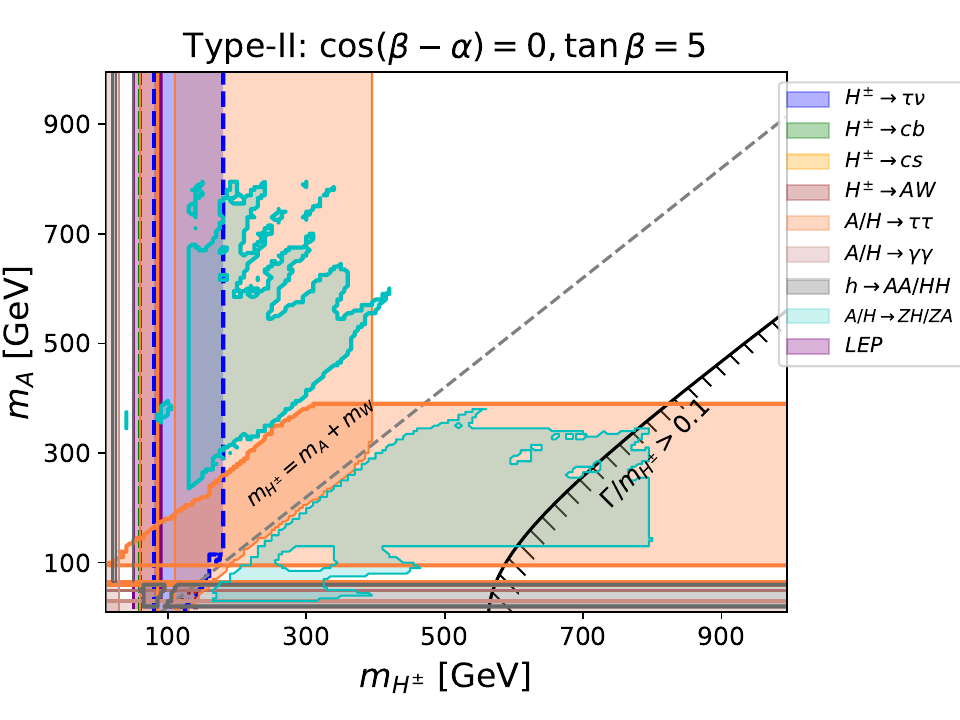}
  \includegraphics[width=0.5\linewidth]{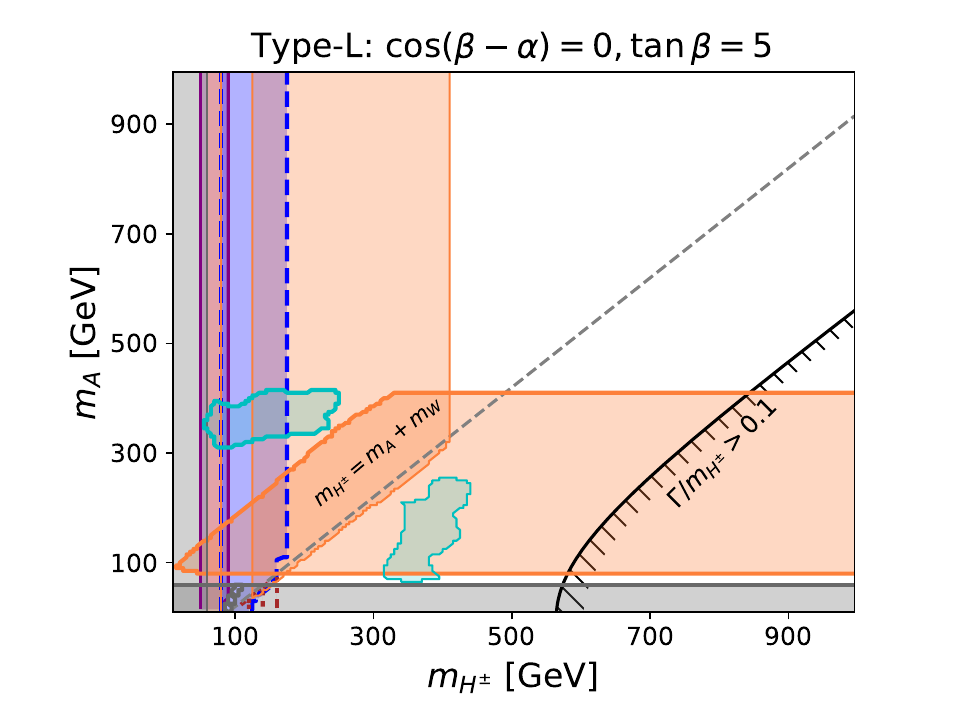}
  \includegraphics[width=0.5\linewidth]{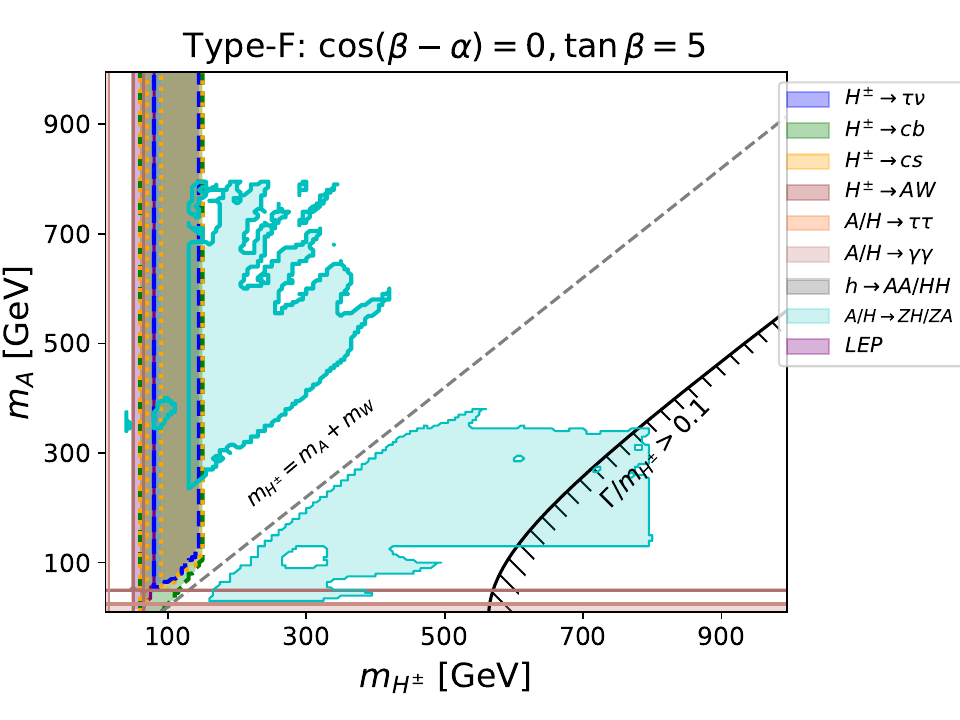}  
  \caption{The 95$\%$ C.L. exclusions region in the $m_{A}$ vs.$m_{H^{\pm }}$ plane in the Type-I 2HDM (upper left), Type-II 2HDM (upper right), Type-L 2HDM (lower left) and Type-F 2HDM (lower right) for $\tan \beta=5$. Color coding is the same as Figure ~\ref{fig:tb1.5}. 
  }
\label{fig:tanb5} 
\end{figure}

 We also show the case of $\tan\beta=5$ in Figure \ref{fig:tanb5}.  Comparing to the $\tan\beta=1.5$ case, the limits get weaker given the reduced production cross section of the charged Higgs.  Only limits from $H^\pm\rightarrow\tau\nu$ (for Type-I, II, L, F), $H^\pm\rightarrow cs$ (for both Type-I and F) and $H^\pm\rightarrow cb$ (Type-F only) survive for light $H^\pm$.  Including the direct search limits for $A$ and $H$, as shown in the bottom panels, provides additional limit for the $m_{A,H}<2m_t$ region.  $A/H \to \tau\tau$ is more effective for the Type-II and Type-L 2HDM, while $A/H \to HZ/AZ$ are more effective in the Type-I, -II, and -F. Note that the excluded region via $A/H \to HZ/AZ$ in the Type-L 2HDM is much smaller, since only a weak constraint from $\tau\tau ll$ final states is applicable here.   Comparing to the $\tan\beta=1.5$ case as shown in~\autoref{fig:tb1.5}, $A/H\to tt$ is absent   since $tt$ channel is suppressed at  $\tan\beta=5$.

\subsubsection{$m_{H^{\pm }}$ vs. $\Delta m_A$}

 \begin{figure}[htbp]
   \includegraphics[width=0.5\linewidth]{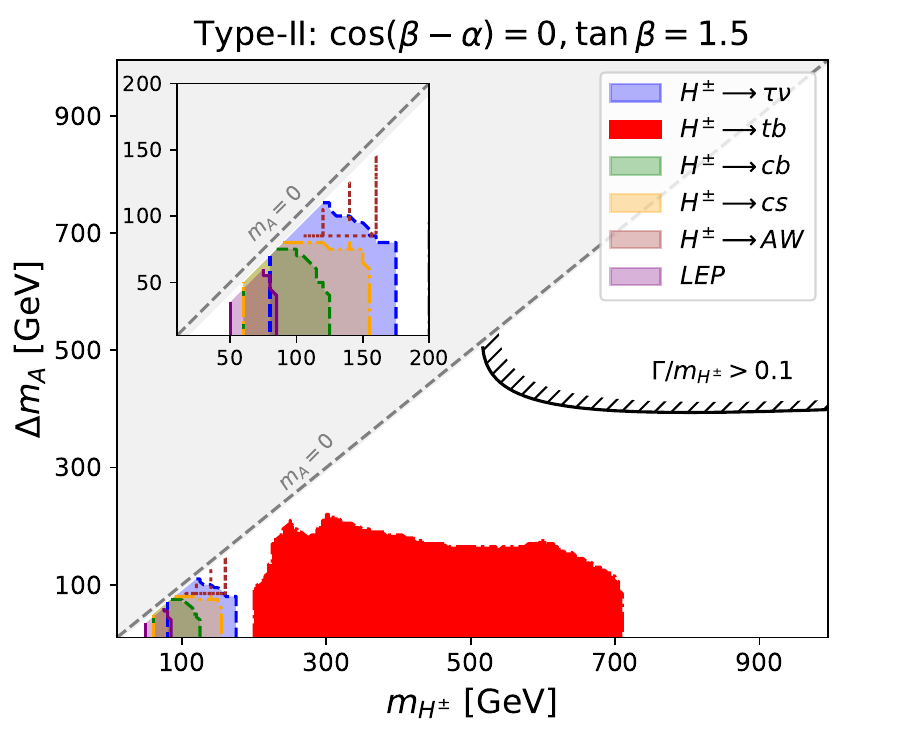}   
   \includegraphics[width=0.5\linewidth]{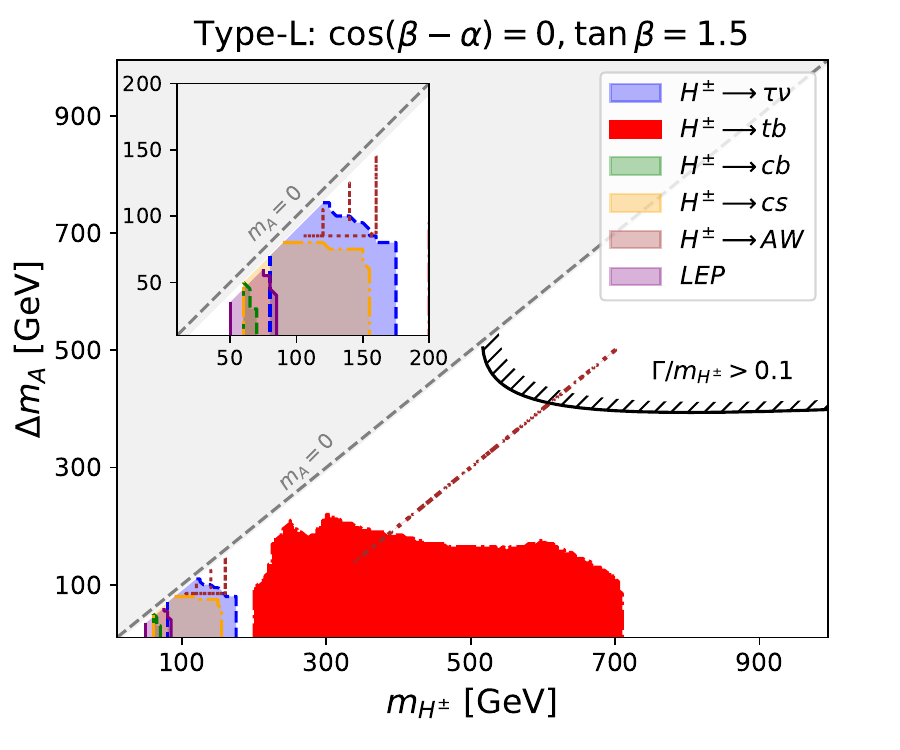}
  \caption{The 95$\%$ C.L. exclusion in the $\Delta m_A=m_{H^{\pm } }- m_{A}$ vs. $m_{H^{\pm } }$ plane of the Type-II 2HDM (left) and Type-L 2HDM (right) for $\tan \beta=1.5$. The upper left corner in each panel shows the zoom-in view of the small mass region. Color coding is the same as Figure~\ref{fig:A85}.}
 \label{fig:mc_ma1.5} 
 \end{figure}

In Figure \ref{fig:mc_ma1.5}, we show the current collider limits in the $m_{H^{\pm } }$ vs. $\Delta m_A$ plane, with $\tan \beta=1.5$ for the Type-II and Type-L 2HDM. The results of Type-I, Type-II, Type-L and Type-F 2HDM are nearly identical for small $\tan\beta$ region, except that the constraint provided by the $H^{\pm } \rightarrow AW$ at heavy $m_{H^\pm}$ is absent in the Type-I, -II, -F.   The top-left half of the plane is not physical since it corresponds to $m_{A}<0$ GeV. The top-right region at $\Delta m_A \gtrsim 400$ GeV corresponding to large decay width: $\Gamma /m_{H^{\pm } } > 0.1$, where the resonant searches of $H^\pm$ are not reliable.  The reduction of the reach at large $m_{H^\pm} \gtrsim 700$ GeV is mainly due to the decrease of the production cross section at large $m_{H^\pm}$.  Note that $H^\pm \rightarrow AW$ channel is indicated by the horizontal $\Delta m_A=85$ GeV line and vertical $m_{H^{\pm }} = 120, 140, 160$ GeV lines in the $m_{H^\pm}<m_t$ region, as well as the $m_A=200$ GeV line parallel to the diagonal line.  
For $\Delta m_A<m_W$, the limits are dominated by the fermionic channels of $\tau \nu$, $cb$, and $cs$ for light $m_{H^\pm}$ and $tb$ for heavy $m_{H^\pm}$.   Once $H^\pm$ decaying to $AW$ is open, the limits from the fermionic channels are relaxed, while $H^{\pm } \rightarrow AW$ provides the best limit for $\Delta m_A\gtrsim 165$ GeV at $\tan\beta=1.5$ in the Type-L 2HDM.

  \begin{figure}[htbp]
   \includegraphics[width=0.5\linewidth]{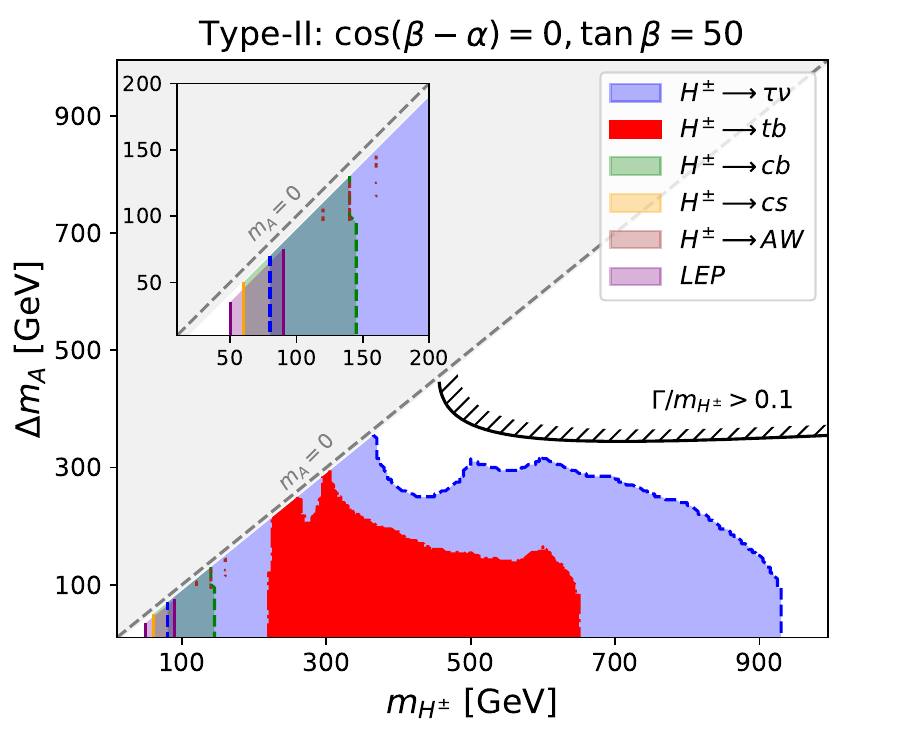}
   \includegraphics[width=0.5\linewidth]{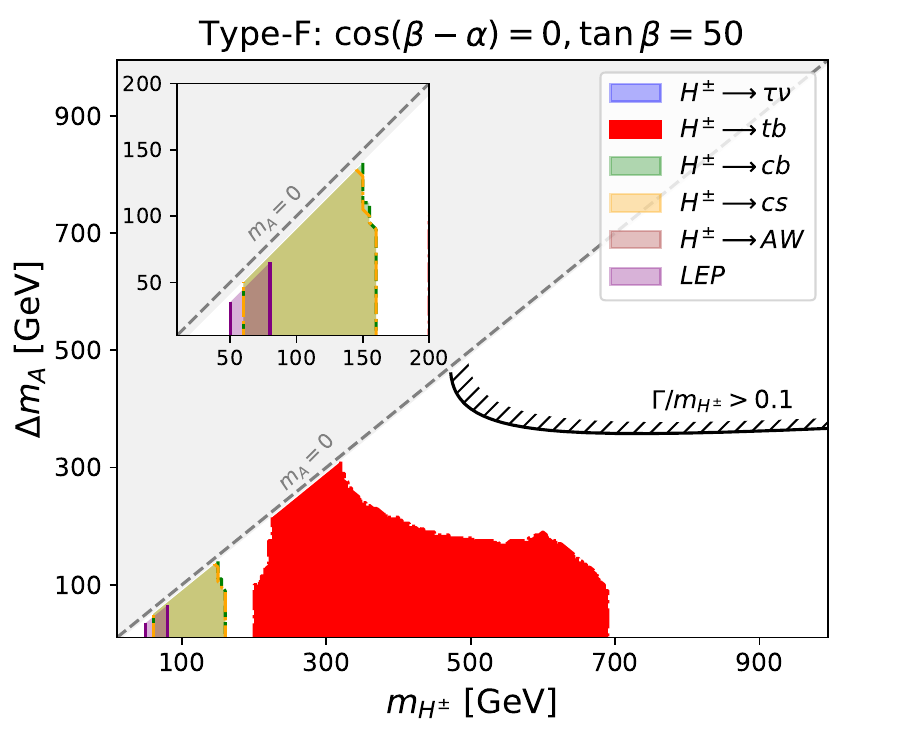}
   \caption{The 95$\%$ C.L. exclusion regions in the $\Delta m_A=m_{H^{\pm } }- m_{A}$ vs.$m_{H^{\pm } }$ plane of the Type-II 2HDM (left) and Type-F 2HDM (right) for $\tan \beta=50$. The upper left corner in each panel shows the zoom-in view of the small mass region. Color coding is the same as Figure~\ref{fig:A85}.}
 \label{fig:mc_ma50} 
 \end{figure}

We also show the case of $\tan \beta =50$ for the Type-II (left) and Type-F (right) in Figure \ref{fig:mc_ma50}. At such large value of $\tan\beta$, there is no constraints on the parameter space of the Type-I and Type-L given the suppressed production cross sections. 
For the Type-II, the parameter space in low mass region is almost entirely excluded by $H^\pm \to \tau\nu$ and  $H^\pm \to cb$ searches. For the heavy $H^\pm$, searches for $H^\pm \to \tau\nu$ and $H^\pm \to tb$ exclude large range of parameter space with $H^\pm \to \tau\nu$ dominating. 
For the Type-F, the parameter space in low mass region is nearly excluded by $H^\pm \to cs$ and  $H^\pm \to cb$ searches. For the heavy $H^\pm$, the exclusion range for $H^\pm \to tb$ is similar to that of the Type-II case, while $H^\pm \to \tau\nu$ exclusion disappears entirely.  Reach of $H^\pm \rightarrow AW$ disappears due to the suppression of the consequent decays of $A \to \mu\mu, \tau\tau$ at such large $\tan\beta$.

\section{Conclusion} \label{sec5}
After the discovery of the SM-like 125 GeV Higgs, the searches for beyond the SM Higgs bosons became an active experimental search frontier. The 2HDM as one of the simplest extensions of the SM predicts the existence of extra Higgs bosons, including a pair of the charged Higgses. In this paper we reinterpreted the current search limits of the charged Higgs boson in the four different types of the 2HDM. We considered constraints from Higgs collider searches at the LEP and the LHC, as well as constraints from flavor physics measurements.  

Collider searches for the charged Higgses mostly focused on the conventional channels of $\tau \nu, cs, cb$, and $tb$. When the exotic decay channels of the charged Higgs boson $H^\pm \rightarrow AW/HW$  are open, the constraints from the fermionic channels are relaxed.  At the same time, $H^\pm \rightarrow AW/HW$ provide alternative search channels for the charged Higgses, which have been performed at the ATLAS and CMS  at 7 TeV, 8 TeV and 13 TeV.  In our analyses, we performed a comprehensive study for both the degenerate spectrum of $m_{H^\pm}=m_H=m_A$ as well as the non-degenerate spectrum of  $m_{H^\pm}=m_H\neq m_A$, when $H^\pm \rightarrow AW$ open.

For the conventional search channels, the most sensitive detection channel for a light charged Higgs with $m_{H^{\pm}} < m_t$ is $H^{\pm} \to \tau \nu$ and $cs$. For the Type-I and Type-L, region of $\tan\beta \lesssim 20$ is excluded. For the high mass region of $m_{H^{\pm}} > m_t$, the most sensitive channel is $H^{\pm} \to tb$ with  $\tan\beta\lesssim 2$  excluding for a charged Higgs mass up to about 1 TeV.  In the Type-II 2HDM,  $H^{\pm} \to \tau \nu$ provides the strongest constraints at the large $\tan\beta$ region. 

When the $H^{\pm} \to AW/HW$ decay channels are kinematically available, the constraints from other conventional search channels are weakened. For $m_{H^\pm} < m_t$,  $H^\pm \to AW$ is complementary to the fermionic channels.
For the heavy charged Higgs $m_{H^\pm} > m_t$, $H^\pm \to tb$ still provides the best search limit, except for the Type-L 2HDM, in which $H^\pm \to AW$ provides the better reach for $1 \lesssim \tan\beta \lesssim 5$.   The current experimental search for the exotic channel of $H^\pm \to AW$  is only performed at a fixed value of $m_A=200$ GeV. To expand the search of this channel over the entire parameter space of $m_{H^\pm}$ vs. $m_A$ could provide a complementary reach to the $H^\pm \to tb$ channel. 

We also studied the limits in the $m_A$ vs. $m_{H^\pm}=m_H$ plane. At small values of $\tan\beta$, almost the entire parameter space up to $m_{H^\pm} \sim 700$ GeV is excluded below the $H^\pm \to AW$ threshold 
with the charged Higgs searches, except a small gap around $m_{H^\pm}\sim m_t$. Once $H^\pm \to AW$ opens, only the reach from the $H^\pm \to tb$ channel extends a little bit beyond the threshold. In the Type-L 2HDM, $H^\pm \to AW$, however, could extend the charged Higgs reach greatly. With the combination of all search channels on both the charged Higgs $H^\pm$ and neutral Higgses $H/A$, almost the entire region of $m_{H^\pm}$ up to about 700 GeV, and $m_A$ up to about 600 GeV is excluded.   The reach, however, is greatly reduced with larger values of $\tan\beta$.

In summary, the exotic charged Higgs search channels $H^\pm \to AW/HW$ are complementary to the conventional fermionic channels. Extending the current $H^\pm \to AW/HW$ searches over the entire $m_H$ vs $m_{A,H}$ plane provides great coverage over the $H^\pm \to tb$ mode.   The combination of all the beyond the SM Higgs searches, including both the neutral and charged ones, could provide crucial insights into the extended Higgs sector.\\ 

\noindent
{\bf Acknowledgments} We would like to thank Adinda De Wit and Alexander Sean Woodcock for insightful discussions.  JL and WS are supported by the Natural Science Foundation of China (NSFC) under grant numbers 12305115 and the Shenzhen Science and Technology Program (Grant No. 202206193000001, 20220816094256002). HS is supported by IBS under the project code, IBS-R018-D1. SS is supported by the Department of Energy under Grant No. DEFG02-13ER41976/DE-SC0009913.


%


%




\newpage
\clearpage
\bibliographystyle{JHEP}
\bibliography{ref_type1}

\end{document}